\documentclass[a4paper,fleqn]{cas-dc}

\usepackage[sort,semicolon]{natbib}

\usepackage[nonumberlist]{glossaries}

\usepackage{url}
\usepackage{xurl} 
\usepackage{makecell} 
\usepackage[british]{babel}
\usepackage[utf8]{inputenc}
\usepackage{textcomp}
\usepackage{gensymb}
\usepackage{color}
\usepackage{subcaption}
\usepackage{float}
\usepackage{stfloats}
\usepackage{units}
\usepackage{mathtools}
\usepackage{amsmath}
\usepackage{cuted}
\usepackage{cleveref}
\usepackage{xcolor}
\usepackage{hyperref}

\newacronym[\glslongpluralkey={slow burn processes}]{SBP}{SBP}{slow burn process}

\newacronym{HILP}{HILP}{high-impact low-probability}
\newacronym{NIAC}{NIAC}{National Infrastructure Advisory Council of the United States}
\newacronym{IEA}{IEA}{International Energy Agency}
\newacronym{ENTSO-E}{ENTSO-E}{European Network of Transmission System Operators for Electricity}
\newacronym{VRE}{VRE}{variable renewable energy}
\newacronym{XB}{XB}{cross-border}
\newacronym{EU}{EU}{European Union}
\newacronym{VOLL}{VOLL}{value of lost load}
\newacronym{LOLE}{LOLE}{loss of load expectation}
\newacronym{EENS}{EENS}{expected energy not served}
\newacronym{SAIDI}{SAIDI}{system average interruption duration index}
\newacronym{HHI}{HHI}{Herfindahl-Hirschman Index}
\newacronym{KPI}{KPI}{key performance indicator}
\newacronym{LNG}{LNG}{liquefied natural gas}
\usepackage{graphicx}
\usepackage{adjustbox}
\usepackage{tabularx}
\usepackage{booktabs}
\usepackage{framed} 
\usepackage{multicol} 
\usepackage{setspace} 
\usepackage{enumitem} 

\usepackage{tikz}
\usetikzlibrary{arrows,decorations.pathmorphing,backgrounds,fit,positioning,shapes.symbols,shapes.geometric,chains,fadings,arrows.meta}

\usepackage{pgfplots}

\pgfdeclareradialshading{myring}{\pgfpointorigin}
{
color(0cm)=(transparent!0);
color(5mm)=(pgftransparent!50);
color(1cm)=(pgftransparent!100)
}
\pgfdeclarefading{ringo}{\pgfuseshading{myring}}

\pgfplotsset{compat=1.18}

\newcolumntype{Z}{>{\raggedright\arraybackslash}p{95pt}}
\newcolumntype{Y}{>{\raggedright\arraybackslash}p{50pt}}

\usepackage{nomencl} 
\makenomenclature
\renewcommand*\nompreamble{\begin{multicols}{2}}
\renewcommand*\nompostamble{\end{multicols}}

\def\tsc#1{\csdef{#1}{\textsc{\lowercase{#1}}\xspace}}
\tsc{WGM}
\tsc{QE}
\tsc{EP}
\tsc{PMS}
\tsc{BEC}
\tsc{DE}

\definecolor{Black}{RGB}{0,0,0}
\definecolor{Orange}{RGB}{230,159,0}
\definecolor{SkyBlue}{RGB}{86,180,233}
\definecolor{BluishGreen}{RGB}{0,158,115}
\definecolor{Yellow}{RGB}{240,228,66}
\definecolor{Blue}{RGB}{0,114,178}
\definecolor{Vermillion}{RGB}{213,94,0}
\definecolor{ReddishPurple}{RGB}{204,121,167}
\definecolor{ukpurple}{RGB}{199,16,92}

\usepackage{todonotes}
\makeatletter
\newcommand*{\hyperlinkcite}[1]{\hyper@link{cite}{cite.#1}}
\makeatother

\begin{document}

\let\WriteBookmarks\relax
\def\floatpagepagefraction{1}
\def\textpagefraction{.001}

\shorttitle{Energy Security and Resilience: Reviewing Concepts and Advancing Planning Perspectives}

\shortauthors{Schmitz, Flachsbarth, Plaga, Braun, Härtel}

\title[mode = title]{Energy Security and Resilience: Reviewing Concepts and Advancing Planning Perspectives for Transforming Integrated Energy Systems}

\author[1,2]{Richard Schmitz}[orcid=0000-0001-8153-3180]
\cormark[1] 
\cortext[cor1]{Corresponding author} 
\ead{richard.schmitz@iee.fraunhofer.de}
\credit{Conceptualisation, Funding acquisition, Methodology, Project administration, Visualisation, Writing – original draft, Writing – review and editing}

\author[3]{Franziska Flachsbarth}[orcid=0000-0002-0813-0358]
\ead{f.flachsbarth@oeko.de}
\credit{Conceptualisation, Funding acquisition, Methodology, Writing – review and editing}

\author[4]{Leonie Sara Plaga}[orcid=0000-0001-6339-6100]
\ead{leonie.plaga@ruhr-uni-bochum.de}
\credit{Conceptualisation, Funding acquisition, Methodology, Writing – review and editing}

\author[1,2]{Martin Braun}[orcid=0000-0003-0857-6760]
\ead{martin.braun@uni-kassel.de}
\credit{Funding acquisition, Supervision, Writing – review and editing}

\author[1,2]{Philipp Härtel}[orcid=0000-0002-9706-1007]
\ead{philipp.haertel@iee.fraunhofer.de}
\credit{Conceptualisation, Funding acquisition, Methodology, Project administration, Supervision, Visualisation, Writing – review and editing}


\affiliation[1]{organization={Fraunhofer IEE, Fraunhofer Institute for Energy Economics and Energy System Technology},
    addressline={Joseph-Beuys-Str. 8, 34117 Kassel},
    country={Germany}}
    
\affiliation[2]{organization={University of Kassel, Department of Sustainable Electrical Energy Systems},
    addressline={Wilhelmshöher Allee 73, 34121 Kassel},
    country={Germany}}

\affiliation[3]{organization={Öko-Institut e.V., Research Division Energy \& Climate},
    addressline={Merzhauser Str. 173, 79100 Freiburg},
    country={Germany}}

\affiliation[4]{organization={Ruhr University Bochum, Chair of Energy Systems and Energy Economics},
    addressline={Universitätsstr. 150, 44801 Bochum},
    country={Germany}}

\begin{abstract}
Recent events, including the pandemic, geopolitical conflicts, supply chain disruptions, and climate change impacts, have exposed the critical need to ensure energy security and resilience in energy systems.
We review existing definitions and interrelations between energy security and resilience, conceptualising these terms in the context of energy system transformations.
We introduce a classification of disturbances into shock events and slow burn processes to highlight key challenges to energy system resilience.
Examples illustrate their distinct impacts on technical, economic, and environmental system performance over time.
We compile relevant recourse options across resilience capacity levels and system planning horizons to address these challenges, emphasising actionable strategies for an increasingly integrated energy system.
Finally, we propose policy recommendations to integrate shock events and slow burn processes into future energy system planning, enabling forward-looking decision-making and system design to analyse and mitigate potential disruptions.
\end{abstract}


\begin{highlights}
\item Review of interrelations between energy security, resilience, and related terms
\item Classification of system disturbances into shock events and slow burn processes
\item Compilation of recourse options to provide strategies for an integrated energy system
\item Clarification of trade-offs between resilience levels and system costs
\item Policy recommendations aiming to integrate disturbances into energy system planning
\end{highlights}

\begin{keywords}
Energy system planning \sep Energy security \sep Resilience \sep Shock events \sep Slow burn processes \sep Recourse options \sep Terminology
\end{keywords}

\maketitle

\begin{table*} 
\begin{framed}
\nomenclature{\acrshort{EENS}}{\acrlong{EENS}}
\nomenclature{\acrshort{ENTSO-E}}{\acrlong{ENTSO-E}}
\nomenclature{\acrshort{EU}}{\acrlong{EU}}
\nomenclature{\acrshort{HHI}}{\acrlong{HHI}}
\nomenclature{\acrshort{HILP}}{\acrlong{HILP}}
\nomenclature{\acrshort{IEA}}{\acrlong{IEA}}
\nomenclature{\acrshort{KPI}}{\acrlong{KPI}}
\nomenclature{\acrshort{LNG}}{\acrlong{LNG}}
\nomenclature{\acrshort{LOLE}}{\acrlong{LOLE}}
\nomenclature{\acrshort{NIAC}}{\acrlong{NIAC}}
\nomenclature{\acrshort{SAIDI}}{\acrlong{SAIDI}}
\nomenclature{\acrshort{SBP}}{\acrlong{SBP}}
\nomenclature{\acrshort{VOLL}}{\acrlong{VOLL}}
\nomenclature{\acrshort{VRE}}{\acrlong{VRE}}
\nomenclature{\acrshort{XB}}{\acrlong{XB}}
\printnomenclature
\end{framed}
\end{table*}

\section{Introduction}
\label{sec:introduction}

Energy security and resilience have become central to national and international policymaking \citep{Siskos.2022}, particularly in the wake of the energy crisis in Europe following the Russia's attack on Ukraine \citep{Giuli.2023}.
While resilience is frequently discussed in engineering, environmental, or social protection contexts \citep{Manca.2017}, the distinction between energy security, resilience, security of supply, and related terms often remains imprecise.
\par
In the past, energy security and security of supply have been evaluated primarily based on the availability and affordability of fossil resources \citep{LeCoq.2009, Vivoda.2009, Cohen.2011, Mansson.2014, Novikau.2023}.
However, in decarbonised energy systems, where fossil fuels play a minimal role, new measures and assessments for evaluating security of supply are required \citep{Kim.2025}.
Recent reviews highlight that security of supply now encompasses dimensions beyond fossil fuel availability \citep{Aisyah.2024, Yu.2022}.
Nevertheless, most studies primarily assess existing systems and rarely account for future uncertainties and risks \citep{Mansson.2014, Gasser.2020b, Axon.2021}.
Moreover, many analyses rely on energy security indices based on factors such as energy availability and pricing \citep{Aisyah.2024}, yet these indices fail to capture the energy system's dynamic nature and adaptive capacity \citep{Martisauskas.2018}.
\par
Resilience research often focuses on the electricity system and its individual components, such as power lines or power plants \citep{Zidane.2025}.
However, as \cite{Jasiunas.2021} argue, the concept of resilience must be applied to the entire integrated energy system in the future.
In long-term energy system planning, optimisation models are commonly used to represent the system's dynamic characteristics \citep{DeCarolis.2017}. 
However, the existing literature provides limited guidance on effectively assessing energy security and resilience within long-term energy system planning.
While several reviews have synthesised existing definitions of energy security, resilience and security of supply \citep{Blum.2012, Bento.2024, Azzuni.2018, Jasiunas.2021}, to our knowledge, no review specifically addresses these terms in the context of long-term energy system optimisation models.
\par
When examining potential threats and hazards to energy security and resilience, a distinction can be made between shock events and so-called \glspl{SBP} \citep{Hanke.2021}.
The negative impact of such events on the energy system is often described as a deterioration in ``energy system performance'' \citep{Braun.2020,Jesse.2019,Jasiunas.2021}, yet this term typically lacks differentiation between technical, economic, or environmental performance.
Our work aims to introduce such a distinction and propose an approach to track performance trends over time.
\par
In considering how energy security and resilience concepts can be applied to planning tasks for future integrated energy systems, a key question arises: what recourse options are available and can be used by decision makers to respond to shock events and \glspl{SBP}?
For instance, energy system planners traditionally rely on historical weather patterns, yet these are becoming increasingly unreliable indicators of future conditions, undermining robust planning for resilience.
We provide an overview of potential recourse options across various planning horizons.
\par
Building on the research gaps identified thus far and aiming to support policymakers in formulating strategies that balance immediate energy needs with long-term sustainability goals, this work makes the following key contributions:
\begin{itemize}
    \item A comprehensive review of previous definitions and a mapping of interrelations between energy security and resilience terms in the energy system domain, incorporating more than 180\,references;
    \item The introduction of shock events and \acrfullpl{SBP} as challenges to energy system resilience, including illustrative examples of their effects on technical, economic, and environmental system performance over time;
    \item A compilation of relevant recourse options at different resilience capacity levels and system planning horizons as responses to disturbances;
    \item A clarification of trade-offs between resilience levels and system costs, recognising the need for political guidelines on desired system resilience levels and acceptable additional costs;
    \item Policy recommendations on how shock events and \glspl{SBP} can be incorporated into the planning task of integrated energy systems to support better-informed, sustainable political decision-making.
\end{itemize}

The remainder is organised as follows.
\Cref{sec:literatureReview} presents an overview of key definitions related to energy security and resilience, examining their interrelations within the literature.
\Cref{sec:shocks_SBPs} defines shock events and \glspl{SBP} as potential challenges to energy security and resilience, followed by illustrative analysis of their impact on the energy system over time.
\Cref{sec:transformation_pathway} explores resilience perspectives in integrated energy system planning and introduces a conceptual approach that categorises relevant recourse options of integrated energy systems with regard to different planning horizons.
\Cref{sec:modelling} outlines the implications of these findings for integrated energy system planning while also hinting at occurring trade-offs between system resilience levels and system costs.
Finally, \Cref{sec:conclusion} discusses policy implications and draws relevant conclusions.
\section{Literature review}
\label{sec:literatureReview}

This literature review aims to provide a comprehensive overview of key definitions and concepts related to energy security and resilience in energy systems.
\Cref{subsec:terms} presents commonly used definitions from the literature, followed by a discussion of their interrelations in \Cref{subsec:interrelations_between_terms}.
Additional terms that contribute to the broader context, namely efficiency, vulnerability, and robustness, are categorised and discussed in \Cref{app:further_terms}.

\subsection{Review of key terms}
\label{subsec:terms}

To establish a comprehensive mapping of interrelations between energy security and resilience terms in the energy system domain, it is first necessary to review relevant terms.

\subsubsection{Energy security}
\label{subsubsec:energy_security}

Although no consensus on a universal definition of energy security exists \citep{Pereira.2024, RodriguezFernandez.2022, Ang.2015, Kim.2025}, it is widely recognised as a key objective of energy policy \citep{Winzer.2012}.
The concept rapidly evolves \citep{Strojny.2023} and closely intersects with energy system resilience \citep{Jasiunas.2021}.
The definition of energy security by the \gls{IEA} as the ``uninterrupted availability of energy sources at an affordable price'' \citep{IEA.2024} features prominently in the academic discourse (e.g. in \citealp{Molyneaux.2016,Pereira.2024,Mauro.2024}).
Another widely referenced framework is the ``four A's'' of energy security, first introduced by \cite{Kruyt.2009} and frequently cited in the literature (e.g. in \citealp{Molyneaux.2016, LaBelle.2024, Strojny.2023, RodriguezFernandez.2022, Cherp.2011, Thaler.2022, Esfahani.2021}).
Availability pertains to physical resources, accessibility involves the geopolitical aspects of acquiring resources, affordability relates to energy costs and prices, and acceptability addresses environmental and social considerations \citep{Kruyt.2009}.
In addition, other dimensions of energy security are highlighted in the literature, including infrastructure \citep{Qiu.2023, Ang.2015}, import dependency and diversity of energy sources \citep{Bento.2024,Mersch.2024}, governance and energy efficiency \citep{Ang.2015, Esfahani.2021, Qiu.2023}, and climate protection \citep{Ang.2015, Buck.2022}.
\par
As noted in the introduction, traditional energy security metrics have historically focused on fossil fuel availability \citep{Kruyt.2009}.
More recent literature identifies indicators such as energy reserves, import exposure, and energy prices as relevant measures of energy security \citep{Zhang.2024}.
Comprehensive tabular overviews of various energy security indicators are available in \cite{Ang.2015,Siksnelyte.2024}.
However, assessing energy security through a single \gls{KPI} remains a challenge due to its multidimensional nature \citep{Ang.2015}.

\subsubsection{Resilience}
\label{subsubsec:resilience}
The concept of resilience has been widely discussed in scientific literature, with early foundational work by \cite{Holling.1973} defining ecosystem resilience as the ability to absorb changes in environmental conditions and parameters.
This definition has since influenced numerous fields (e.g. in \citealp{Wang.2019,Hanke.2021,Brand.2016}).
The term itself is derived from the Latin ``resilire'', meaning ``the ability to spring back or rebound'' \citep{Gholami.2018,Gatto.2020b,Hosseini.2016}.
\par
Despite its widespread use, no single, universally accepted definition of resilience exists, as its meaning varies significantly across disciplines and contexts \citep{Wang.2019,Jesse.2024,Erker.2017b,Mentges.2023,Gasser.2020}.
It is often described as a multidimensional concept \citep{Altherr.2018,Manca.2017,Siskos.2022}, encompassing diverse aspects such as energy import resilience \citep{He.2015}, resilient supply chains \citep{Balteanu.2024}, resilient energy communities \citep{Gruber.2024}, and resilience of critical infrastructures \citep{Mentges.2023}.
From a general perspective, \cite{Erker.2017} define resilience as ``a concept to ensure the viability of a system by reducing the vulnerability and increasing the adaptive capacity before, during or after a stressful event''.
Similar definitions appear in \cite{Cho.2022}, \cite{Jasiunas.2021}, \cite{VDE.2020}, and \cite{Zhou.2023}.
\par
Energy system resilience is defined as ``the ability of an energy system to retain, react, overcome and overpass perturbations caused by a shock in economic, social, environmental and institutional terms'' \citep{Gatto.2020,Ye.2024}.
It is also important to note the definitions of relevant institutions:
The \cite{EU.2022} frames resilience of a critical organisation or infrastructure as the ability to prevent, protect against, respond to, mitigate the consequences of, absorb, manage, and recover from a security incident.
In turn, the \gls{ENTSO-E} defines power system resilience as ``the ability to withstand and mitigate the extent, severity and duration of system degradation following an impactful event'' \citep{ENTSOE.2024}.
Almost identical, the International Council on Large Electric \mbox{Systems (CIGRE)} defines power system resilience as ``the ability to limit the extent, severity and duration of system degradation following an extreme event'' \citep{CIGRE.2017}.
Similar adaptions of these definitions can be found in \cite{Braun.2023}, \cite{Izadi.2021}, \cite{Pitto.2024}, and \cite{Stankovic.2023}.
\par
To characterise resilience, the \gls{NIAC} identifies robustness, resourcefulness, rapid recovery, and adaptability as its key features \citep{NIAC.2010}.
According to the \cite{CabinetOffice.2011}, components of resilient infrastructure are resistance, redundancy, reliability, response, and recovery.
\par
Resilience is closely linked to \gls{HILP} events \citep{Janta.2024,Braun.2020,Stankovic.2023,Zhou.2020,Cho.2022,Mohanty.2024,Zou.2024,Pan.2025,Fang.2019} that are sometimes also referred to as extreme events \citep{Broska.2020}, disruptive events \citep{Mentges.2023}, or black swan hazards \citep{Panteli.2018,Krupa.2013}.
A resilience trapezoid is commonly employed to track the (energy) system's performance over time to visually represent the resilience of an (energy) system in case of such an \gls{HILP} event (e.g. in \citealp{Amini.2023,Braun.2020,Izadi.2021,Janta.2024,Jasiunas.2021,Jesse.2019,Mohanty.2024,Panteli.2017,Panteli.2017b,Stankovic.2023}).
Although terminology varies across sources, the different phases of resilience assessment typically fall under the categories of defending, preventing, detecting, resisting, absorbing, adapting, and restoring.
\par
The assessment of energy system resilience often considers both the duration and severity of disruptions \citep{Panteli.2017}.
Another key metric is the elapsed time from the occurrence of a disruption to full system recovery.
However, relying solely on a single or aggregated energy resilience \gls{KPI} may not provide a complete picture.
Instead, researchers advocate for a multi-indicator framework to capture the complexity of resilience assessment.
Comprehensive tabular overviews of various resilience indicators are available in \cite{Ahmadi.2021}, \cite{Shafiei.2025}, and \cite{Monie.2025}.

\subsubsection{Energy sovereignty}
\label{subsubsec:energy_sovereignty}

Energy sovereignty has gained significant attention in recent years \citep{Prognos.2023,Timmermann.2022}.
As \cite{Raimi.2024} note, a universally accepted definition remains elusive.
However, they characterise energy sovereignty as a framework that enables communities or nations to make informed choices regarding all components of their energy systems.
\cite{LaBelle.2024} further defines energy sovereignty as a state's capability to safeguard its energy system against multidimensional threats through control over its energy policy.
Similarly, \cite{Westphal.2020} argues that energy sovereignty is achieved when a state can provide sufficient and reliable energy supplies at economic prices without compromising its own values, interests, or foreign policy objectives.
\par
The Observatori del Deute en la Globalitzaci{\'o} \citep{ODG.2014} defines energy sovereignty as the right to make independent decisions on energy generation, distribution, and consumption while considering economic, social, and ecological factors.
\cite{Thaler.2022} distinguish between `hard' aspects, i.e. independent decisions regarding the structure and sources of energy supply, and `soft' aspects of energy sovereignty, i.e. control over the operation of the energy system, energy policy, and market regulations.
\par
Several \glspl{KPI} have been proposed to assess energy sovereignty, including the share of domestic energy production in total energy consumption, the ratio of domestic renewable energy production to total energy demand, or the net energy import dependency.
Another \gls{KPI} for energy sovereignty is the \gls{HHI} \citep{Bucciarelli.2025}, which measures the concentration of energy supply or production within a particular market or region.
Closely related to energy sovereignty is the concept of technology sovereignty \citep{Edler.2023}, which focuses on the state's ability to produce or control strategic energy technologies domestically.
This can be quantified, for example, as the proportion of strategic energy technologies manufactured within a country.

\subsubsection{Energy solidarity}
\label{subsubsec:energy_solidarity}

The objective of energy solidarity is to ensure the efficient operation of energy markets and the security of energy supply, while promoting energy efficiency and innovation, and enhancing the interconnection of energy networks \citep{Huhta.2023b}.
More broadly, solidarity is a foundational principle of the \gls{EU} \citep{Misik.2023}, implying that Member States are expected to assist one another across various domains, even at the expense of their own national interests \citep{Sangiovanni.2013}.
When applied to the energy sector, this principle is embodied in energy solidarity, which serves as a legally binding principle within the EU's energy law \citep{Huhta.2023}. 
Although no universally accepted definition of energy solidarity exists, there is a general consensus that fostering a spirit of solidarity among Member States is essential for the effective functioning of the energy market \citep{Huhta.2023,Bartenstein.2023,Andoura.2023}.
Based on \cite{EU.2008}, energy solidarity can be defined as the collaborative commitment among countries to ensure a secure, efficient, and sustainable energy market.
\par
The literature provides limited guidance on \glspl{KPI} for measuring energy solidarity.
However, potential metrics could include the volume of cross-border electricity and gas exchange, the share of energy imports from allied countries, or the response time for energy crisis interventions.

\subsubsection{Security of supply}
\label{subsubsec:security_of_supply}

There are various definitions of security of supply \citep{Burkhardt.2024}, which is sometimes considered a dimension of energy security \citep{Fouladvand.2024, Kim.2025}.
However, the literature frequently blurs the distinction between the terms energy security and security of supply \citep{Wettingfeld.2024}.
According to \cite{Most.2023}, security of supply is achieved when the desired quantity of energy (of the required quality) is available consistently throughout the energy system at reasonable prices, which is similar to the definition of energy security.
Security of supply primarily focuses on the reliability of power supply \citep{Loschel.2023} and the adequacy of generation and power grids \citep{Wettingfeld.2024}.
It is often illustrated through the assumption of overcapacities in electricity generation and transmission \citep{Gils.2023}.
According to the German Federal Network Agency (\citealp{Bundesnetzagentur.2024}), security of supply is characterised by sufficient generating capacities, electricity grid and gas networks that fulfil their transport tasks, reliable control mechanisms, and sufficient protection against third-party interference.
The German Federal Ministry for Economic Affairs and Climate Action \citep{BMWK.2019} defines security of supply as maintaining electricity supply at all times despite an advancing energy transition.
\cite{Schittekatte.2021} categorise security of supply along the temporal dimension into a real-time to short-term aspect (system reliability) and a mid-term to long-term aspect (resource adequacy).
\par
Reliability is viewed as a precursor to resilience \citep{Cramton.2023}, particularly in the context of expansion planning \citep{Espinoza.2016,Cho.2022}.
With regard to power systems, reliability refers to the probability of satisfying the load demand under uncertain conditions \citep{Tsao.2020,Lin.2012}, especially in regionally restricted disruptions.
The \cite{CabinetOffice.2011} explicitly emphasises the design of power system components, defining reliability as the assurance that these components are inherently capable of functioning under various conditions.
Unlike resilience, which deals with \gls{HILP} events, reliability addresses high-probability and low-impact events \citep{Izadi.2021,Braun.2020,Mohanty.2024}.
\par
The term adequacy, commonly referred to as ``system adequacy'', is used in various contexts \citep{Consentec.2015}.
A widely accepted definition by \citep{TERNA.2024} describes it as ``the system’s capacity to satisfy electrical energy requirements while complying with requirements on safety and quality of service''.
This definition mainly encompasses existing grid and generation capacities \citep{Wettingfeld.2024,Consentec.2015}.
The anticipated retirement or phasing out of conventional power plants and the expansion of renewable energy are expected to pose challenges for maintaining generation adequacy in the European power system \citep{Scharf.2024}.
A prominent source monitoring Europe's security of supply over a 10-year horizon is the ``European resource adequacy assessment'', published by the \cite{EuropeanUnionAgency.2024}.
\par
Grid adequacy (or transmission adequacy) is a subset of system adequacy and refers to the ability to transport electricity from generation sites to consumers at all times, which requires sufficient transmission and distribution grid capacity \citep{TenneT.2024}.
\par
Several \glspl{KPI} are commonly used to assess security of supply, including the \gls{LOLE}, \gls{EENS}, \gls{SAIDI}, and \gls{VOLL}.
The \gls{LOLE} measures the expected number of hours per year in which the demand for (electrical) energy cannot be met \citep{Grant.2024}.
The \gls{EENS} quantifies the expected amount of energy that will not be provided due to supply interruptions \citep{Qawaqzeh.2023}.
The \gls{SAIDI} calculates the average duration of power interruptions per customer served over a given period \citep{Bundesnetzagentur.2025}.
The \gls{VOLL} estimates the economic impact of supply interruptions \citep{Kachirayil.2025}, reflecting society's willingness to pay to prevent power outages \citep{Gorman.2022}. The calculation of the \gls{VOLL} is influenced by several factors, including the timing and duration of the outage, as well as regional economic conditions \citep{Najafi.2021}.

\subsection{Interrelations between defined terms}
\label{subsec:interrelations_between_terms}

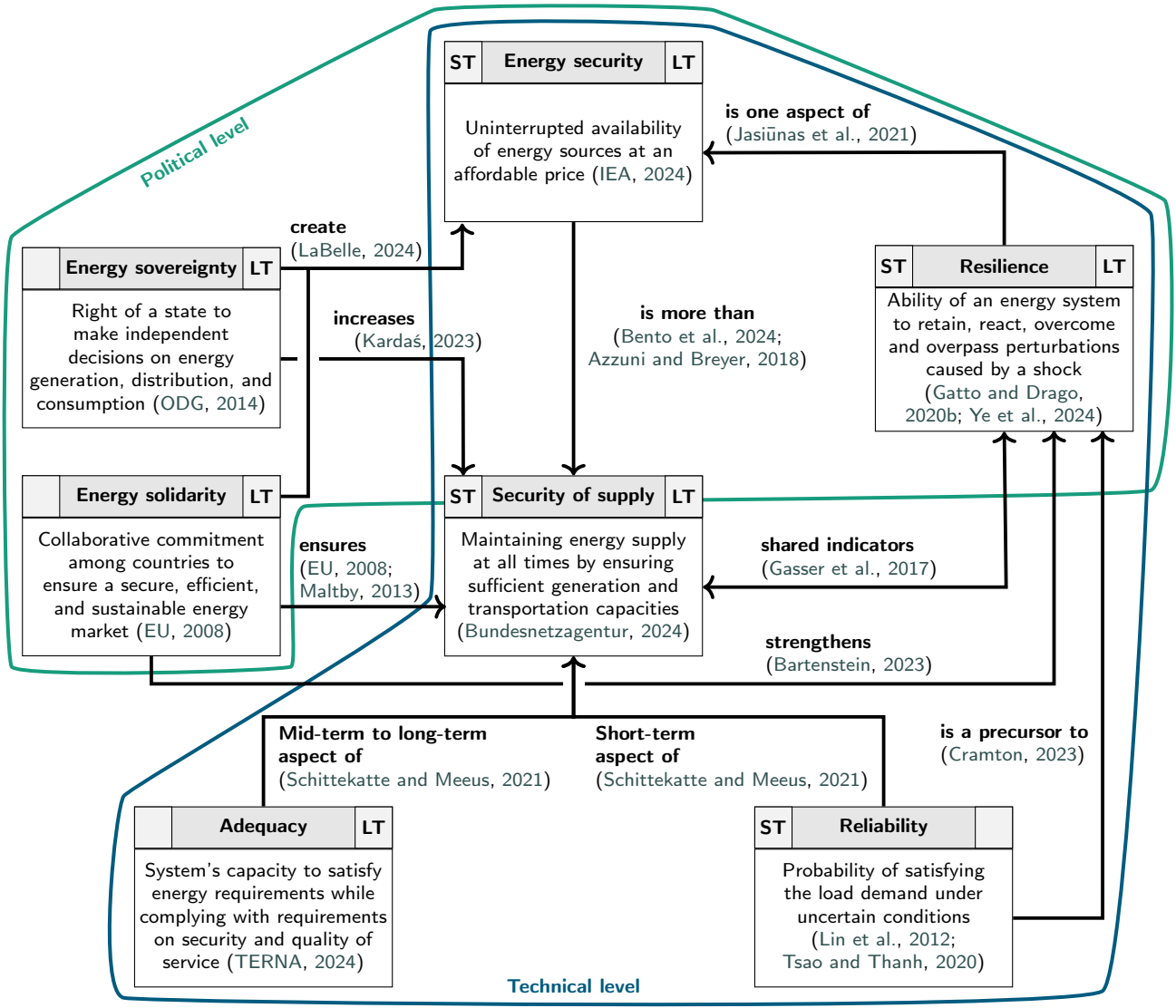
\begin{figure*}[h]
    \centering
    \begin{tikzpicture}
[auto,font=\footnotesize,
every node/.style={node distance=0cm},
category/.style={rectangle, draw, fill=black!10, inner sep=2pt, text width=3.6cm, align=center, minimum height=0.6cm, font=\bfseries\footnotesize},
desc/.style={rectangle, draw, fill=white, inner sep=2pt, text width=3.6cm, align=center, minimum height=2.0cm, font=\footnotesize},
stlt/.style={rectangle, draw, fill=black!5, inner sep=2pt, text width=0.4cm, align=center, minimum height=0.6cm, font=\bfseries\footnotesize}]

\definecolor{fhgreen}{RGB}{23,156,125}
\definecolor{fhblue}{RGB}{0,91,127}
\definecolor{fhgrey}{RGB}{166,187,200}
\definecolor{fhorange}{RGB}{245,130,32}
\definecolor{fhpurple}{RGB}{124,21,77}
\definecolor{lightgrey}{RGB}{196,210,218}

\node [category, anchor=mid] (ensec) {Energy security};
\node [stlt, right=of ensec.north west, anchor=north west] (ensec_st) {ST};
\node [stlt, left=of ensec.north east, anchor=north east] (ensec_lt) {LT};
\node [desc, below=of ensec.south, anchor=north] (ensec_desc) {Uninterrupted availability of energy sources at an affordable price \citep{IEA.2024}};

\node [category, below=6.0cm of ensec, anchor=mid] (secsup) {Security of supply};
\node [stlt, right=of secsup.north west, anchor=north west] (secsup_st) {ST};
\node [stlt, left=of secsup.north east, anchor=north east] (secsup_lt) {LT};
\node [desc, below=of secsup.south, anchor=north] (secsup_desc) {Maintaining energy supply at all times by ensuring sufficient generation and transportation capacities \citep{Bundesnetzagentur.2024}};

\node [category, left=4.25cm of secsup, anchor=mid] (ensol) {Energy solidarity};
\node [stlt, right=of ensol.north west, anchor=north west] (ensol_st) {};
\node [stlt, left=of ensol.north east, anchor=north east] (ensol_lt) {LT};
\node [desc, below=of ensol.south, anchor=north] (ensol_desc) {Collaborative commitment among countries to ensure a secure, efficient, and sustainable energy market \citep{EU.2008}};

\node [category, above=3.0cm of ensol, anchor=mid] (ensov) {Energy sovereignty};
\node [stlt, right=of ensov.north west, anchor=north west] (ensov_st) {};
\node [stlt, left=of ensov.north east, anchor=north east] (ensov_lt) {LT};
\node [desc, below=of ensov.south, anchor=north] (ensov_desc) {Right of a state to make independent decisions on energy generation, distribution, and consumption \citep{ODG.2014}};

\node [category, right=10.5cm of ensov, anchor=mid] (resi) {Resilience};
\node [stlt, right=of resi.north west, anchor=north west] (resi_st) {ST};
\node [stlt, left=of resi.north east, anchor=north east] (resi_lt) {LT};
\node [desc, below=of resi.south, anchor=north] (resi_desc) {Ability of an energy system to retain, react, overcome and overpass perturbations caused by a shock\\\citep{Gatto.2020,Ye.2024}};

\node [category, below=4.5cm of secsup, anchor=mid, xshift=-4.5cm] (adeq) {Adequacy};
\node [stlt, right=of adeq.north west, anchor=north west] (adeq_st) {};
\node [stlt, left=of adeq.north east, anchor=north east] (adeq_lt) {LT};
\node [desc, below=of adeq.south, anchor=north] (adeq_desc) {System's capacity to satisfy energy requirements while complying with requirements on security and quality of service \citep{TERNA.2024}};

\node [category, below=4.5cm of secsup, anchor=mid, xshift=+4.5cm] (relia) {Reliability};
\node [stlt, right=of relia.north west, anchor=north west] (relia_st) {ST};
\node [stlt, left=of relia.north east, anchor=north east] (relia_lt) {};
\node [desc, below=of relia.south, anchor=north] (relia_desc) {Probability of satisfying the load demand under uncertain conditions\\(\citealp{Lin.2012};\\\citealp{Tsao.2020})};

\draw[<->, line width=0.5mm] (secsup_desc.east) --++ (4.40,0) -- node[left, above, sloped]{} (resi_desc.south); 

\draw[->, line width=0.5mm] (resi.north) --++ (0,1.35) -- node[left, above, sloped]{} (ensec_desc.east); 

\draw[->, line width=0.5mm] (ensec_desc.south) -- node[left, above, sloped]{}(secsup.north); 

\draw[->, line width=0.5mm] (ensov_desc.east) --++ (2.645,0) -- node[left, above, sloped]{}(secsup_st.north); 

\fill[white] (-4.005, -4.37) rectangle (-3.695, -4.24); 

\draw[->, line width=0.5mm] (ensov_lt.east) --++ (2.635,0) --++ (0,0.65); 

\draw[-, line width=0.5mm] (ensol_lt.east) --++ (0.4,0) --++ (0,3.3); 

\draw[->, line width=0.5mm] (relia_desc.east) --++ (1.3,0) --++ (0,7.07); 

\draw[->, line width=0.5mm] (ensol_desc.south) --++ (0,-0.4) --++ (13.1,0) --++ (0,3.67); 

\draw[->, line width=0.5mm] (-4.25,-7.9) --++ (2.40,0); 

\fill[white] (-0.155, -9.08) rectangle (0.155, -8.95); 

\draw[->, line width=0.5mm] (adeq.north) --++ (0,1.3) --++ (4.5,0) -- node[left, above, sloped]{}(secsup_desc.south); 

\draw[->, line width=0.5mm] (relia.north) --++ (0,1.3) --++ (-4.5,0) -- node[left, above, sloped]{}(secsup_desc.south); 

\node[align=center] at (1.8,-4.0) {\textbf{is more than}\\(\citealp{Bento.2024};\\\citealp{Azzuni.2018})};

\node[align=left] at (3.6,-0.9) {\textbf{is one aspect of}\\\citep{Jasiunas.2021}};

\node[align=left] at (-3.15,-2.6) {\textbf{create}\\\citep{LaBelle.2024}};

\node[align=left] at (-2.4,-3.9) {\textbf{increases}\\\quad \citep{Kardas.2023}};

\node[align=left] at (-3.1,-7.4) {\textbf{ensures}\\(\citealp{EU.2008};\\\citealp{Maltby.2013})};

\node[align=left] at (4.0,-7.2) {\textbf{shared indicators}\\\citep{Gasser.2017}};

\node[align=left] at (4.0,-8.6) {\textbf{strengthens}\\\citep{Bartenstein.2023}};

\node[align=left] at (-2.3,-10.1) {\textbf{Mid-term to long-term}\\\textbf{aspect of}\\\citep{Schittekatte.2021}};

\node[align=left] at (2.3,-10.1) {\textbf{Short-term}\\\textbf{aspect of}\\\citep{Schittekatte.2021}};

\node[align=left] at (6.4,-9.9) {\textbf{is a precursor to}\\\citep{Cramton.2023}};

\begin{scope}[on background layer]

\draw[fhgreen,draw opacity=1,line width=1.5pt] plot [smooth cycle, tension=0.1] coordinates {
([xshift=+0.45cm,yshift=+0.65cm]resi.north east)
([xshift=+0.25cm,yshift=-0.85cm]resi_desc.south east)
([xshift=+0.35cm,yshift=-0.45cm]ensol.north east)
([xshift=+0.15cm,yshift=-0.15cm]ensol_desc.south east)
([xshift=-0.15cm,yshift=-0.15cm]ensol_desc.south west)([xshift=-0.15cm,yshift=0.15cm]ensov.north west)
([xshift=-0.15cm,yshift=0.45cm]ensec.north west)
([xshift=+0.15cm,yshift=0.45cm]ensec.north east)
};

\draw[fhblue,draw opacity=1,line width=1.5pt] plot [smooth cycle, tension=0.1] coordinates {
([xshift=+0.25cm,yshift=0.45cm]resi.north east)
([xshift=+1.30cm,yshift=-0.05cm]relia_desc.south east)
([xshift=-0.15cm,yshift=-0.15cm]adeq_desc.south west)
([xshift=-0.15cm,yshift=+0.15cm]adeq.north west)
([xshift=-0.20cm,yshift=-0.15cm]secsup_desc.west)
([xshift=-0.20cm,yshift=+0.15cm]ensec.north west)
([xshift=+0.15cm,yshift=+0.20cm]ensec.north east)
};

\end{scope}

\node[text=fhgreen, rotate=30] at (-5.5,-1.3) {\textbf{Political level}};
\node[text=fhblue, rotate=0] at (0,-13.4) {\textbf{Technical level}};

\end{tikzpicture}
    \caption{Interrelations between definitions for terms relevant to energy security and resilience in energy systems and their transformation, own illustration. The white text boxes display commonly cited definitions from the literature. The labels ``ST'' and ``LT'' indicate whether terms are typically associated with a short-term or long-term context. Directed arrows represent the interrelations between terms.}
    \label{fig_mindmap}
    \vspace{-1.0em}
\end{figure*}

Based on the literature review, a conceptual framework illustrating the definitions and interrelations between key terms relevant to energy security and resilience in energy systems has been developed (see \Cref{fig_mindmap}).
The overlapping arrows in the figure indicate that these interrelations do not follow a simple linear hierarchy but are instead strongly interlinked.

Starting from the left side, \cite{LaBelle.2024} argues that energy sovereignty and energy solidarity contribute to the creation of energy security rather than being variables independent from energy security.
Both energy sovereignty and energy solidarity play a supportive role in enhancing security of supply: the former increases security of supply, according to \cite{Kardas.2023}, while the latter ensures the security of energy supply in the \gls{EU}, according to \cite{EU.2008} and \cite{Maltby.2013}.
Furthermore, \cite{Bartenstein.2023} suggests that energy solidarity also strengthens the resilience of the (European) energy system.
Resilience is often regarded as a component of energy security \citep{Jasiunas.2021}.
According to \citep{Gasser.2017}, many indicators of resilience can also be applied in the context of security of supply.
\cite{Bento.2024} and \cite{Azzuni.2018} both argue that energy security is more than just security of supply because of other economic, social, and environmental aspects that also belong to the field of energy security.
As discussed in \Cref{subsubsec:security_of_supply}, and according to \cite{Schittekatte.2021}, reliability represents the real-time to short-term aspects of security of supply, while adequacy pertains to the mid-term to long-term elements of security of supply.
\par
\Cref{fig_mindmap} also categorises the different terms based on their association with either the technical or political level.
Energy security and resilience appear across both technical and political levels, encompassing infrastructure, market dynamics, and policy considerations.
Energy sovereignty and energy solidarity are primarily linked to the political level, given their association with policy decisions, international cooperation, and regulatory frameworks.
Security of supply, along with reliability and adequacy, is primarily considered a technical concept.
However, it also borders the political level, as decisions on market design, capacity mechanisms, and cross-border cooperation influence its effectiveness.
\section{Shocks and SBPs challenging resilience}
\label{sec:shocks_SBPs}

The interrelations between the terms discussed in \Cref{sec:literatureReview} demonstrate that enhancing the energy system's resilience contributes inherently to improvements in both energy security and security of supply.
Consequently, the primary focus of this work is to explore strategies for strengthening energy system resilience, recognising that such efforts will also support energy security and security of supply. 
For the sake of simplicity, however, the following sections will primarily refer to energy system resilience.
\par
As identified in the literature review, resilience in energy systems is particularly challenged by hazards and threats.
While hazards typically refer to natural and unintentional events, threats are linked to intentional actions driven by specific actors with the capability and intent to cause harm \citep{Mentges.2023}.
\par
In the context of energy system planning and modelling, such events can be classified as either transient or disruptive.
Transient events are anticipated to some extent but may not be fully accounted for in planning.
Disruptive events arise from large-scale mega-trends with unpredictable magnitudes and progression speeds \citep{McCollum.2020}.
Note that in a narrower context of electrical engineering, transient events are defined as temporary, short-lived phenomena.
In macro-scale energy system modelling, the term is used more broadly to describe events with similar characteristics but on a different timescale.
\par
In order to optimise system design and enhance resilience, it is essential to consider hazards and threats in planning future energy systems. 
\cite{Hanke.2021} introduce a useful distinction between ``shock events'' and so-called \acrfullpl{SBP}, which provides a structured approach to classifying hazards and threats as well as examining their effects on energy systems' technical, economic, and environmental performances.
\Cref{subsec:def_shock_SBP} first characterises the two terms, followed by a discussion of their impact over time in \Cref{subsec:impact_shock_SBP_over_time}.

\subsection{Characteristics of shock events and SBPs}
\label{subsec:def_shock_SBP}

As discussed in \Cref{subsubsec:resilience}, shock events are often described as \gls{HILP} events.
Although such events are rare, their potential for substantial and difficult-to-mitigate damage is significant.
Importantly, shock events cannot be entirely avoided \citep{Manca.2017}.
Their impact is typically unevenly distributed across countries \citep{Zachmann.2024}.
In highly interconnected energy systems, such disruptions can propagate across neighbouring regions amplifying their effects.
\par
In contrast, \acrfullpl{SBP} refer to protracted, persistent, structural changes in the energy system \citep{Manca.2017,Hanke.2021}.
They evolve gradually and escalate in urgency over time.
While unfolding, they lead to increased vulnerability and diminishing resilience, potentially resulting in a shock event in the future.
A frequently used example is climate change, which will have a profound impact on the energy system due to increased extreme weather events, global warming, and changing resource requirements.
\par
Various hazards and threats that pose potential risks to the energy system have been identified in the literature.
A classification of these risks into shock events and \glspl{SBP} is provided in \Cref{tab_shocks}.
Beyond commonly discussed shock events in the table, the literature also identifies less frequently explored risks, including individual but more pronounced droughts of solar and wind \citep{Kapica.2024,Grochowicz.2024}, or pandemics \citep{Hoang.2021}.
Note that it is crucial to distinguish between their respective incidents and impacts \citep{Broska.2020}.
Therefore, the table focuses on event classification rather than detailing their consequences for the energy system.

\begin{table*}[h]
\begin{center}
\caption{Overview of potential hazards and threats to the energy system including a classification into shock events and \glspl{SBP}, with references from the literature.}
\label{tab_shocks}
\begin{tabularx}{490pt}{ZYX} 
\toprule
\textbf{Potential hazard \newline or threat} & \textbf{Shock event \newline or \gls{SBP}} & \textbf{References}\\
\midrule
Hurricanes, Tornadoes, Storms & Shock event & \cite{Amini.2023}, \cite{Braun.2023}, \cite{CharaniShandiz.2020}, \cite{Espinoza.2016}, \cite{Fang.2019}, \cite{Janta.2024}, \cite{Jasiunas.2021}, \cite{Lau.2023}, \cite{Manca.2017}, \cite{Mohanty.2024}, \cite{Most.2023}, \cite{Osman.2023}, \cite{Panteli.2017b}, \cite{Roege.2014}, \cite{Stankovic.2023}, \cite{Wang.2019}, \cite{Yang.2024}, \cite{Zhou.2020}\\
\midrule
Flood, Heavy rain & Shock event & \cite{Amini.2023}, \cite{Braun.2023}, \cite{CharaniShandiz.2020}, \cite{Espinoza.2016}, \cite{Fang.2019}, \cite{Hawker.2024}, \cite{Janta.2024}, \cite{Jasiunas.2021}, \cite{Lau.2023}, \cite{Manca.2017}, \cite{Martisauskas.2018}, \cite{Mohanty.2024}, \cite{Most.2023}, \cite{Panteli.2017b}, \cite{Stankovic.2023}, \cite{Yang.2024}, \cite{Zhou.2020}\\
\midrule
Tsunami, Earthquake, Volcanic activity & Shock event & \cite{Amini.2023}, \cite{Braun.2023}, \cite{CharaniShandiz.2020}, \cite{Espinoza.2016}, \cite{Fang.2019}, \cite{Manca.2017}, \cite{Mohanty.2024}, \cite{Roege.2014}, \cite{Underwood.2020}\\
\midrule
Extreme temperature, Drought, Heat waves, Forest fire & Shock event & \cite{Abdin.2019}, \cite{CharaniShandiz.2020}, \cite{Fang.2019}, \cite{Gils.2023}, \cite{Hawker.2024}, \cite{Janta.2024}, \cite{Jasiunas.2021}, \cite{Lau.2023}, \cite{Manca.2017}, \cite{Martisauskas.2018}, \cite{Mohanty.2024}, \cite{Osman.2023}, \cite{Stankovic.2023}, \cite{vanderMost.2024}, \cite{Yang.2024}, \cite{Zhou.2020}\\
\midrule
Extreme cold, Snow storms, Blizzards & Shock event & \cite{Cramton.2023}, \cite{Espinoza.2016}, \cite{Fang.2019}, \cite{Gils.2023}, \cite{Hawker.2024}, \cite{Most.2023}, \cite{Panteli.2017b}, \cite{Stankovic.2023}, \cite{Zhou.2020}\\
\midrule
Space weather & Shock event & \cite{Amini.2023}, \cite{Braun.2023}, \cite{Jasiunas.2021}, \cite{Liu.2024}, \cite{Taran.2023}\\
\midrule
Terrorist attacks & Shock event & \cite{Braun.2024}, \cite{CharaniShandiz.2020}, \cite{Jesse.2019}, \cite{Lee.2022}, \cite{Martisauskas.2018}, \cite{Most.2023}, \cite{Wang.2019}\\
\midrule
Cyber-physical attacks & Shock event & \cite{Amini.2023}, \cite{Braun.2023}, \cite{Diaba.2024}, \cite{Gils.2023}, \cite{Martisauskas.2018}, \cite{Stankovic.2023}, \cite{Zhao.2024}\\
\midrule
Climate change & \gls{SBP} & \cite{Bento.2024}, \cite{Braun.2024}, \cite{Craig.2022}, \cite{Cronin.2018}, \cite{Gils.2023}, \cite{Guenand.2024}, \cite{Jesse.2019}, \cite{Plaga.2023}, \cite{Sundar.2024}\\
\midrule
``Dunkelflaute'' & \gls{SBP} & \cite{ENTSOE.2021}, \cite{Gils.2023}, \cite{Grochowicz.2024}, \cite{Kittel.2024}, \cite{Mayer.2023}, \cite{Ohba.2022}, \cite{vanderMost.2024}\\
\midrule
Uncontrollable excess electricity (``Helle Brise'' or ``Warmer Lichtsturm'') & \gls{SBP} & \cite{50Hertz.2024}, \cite{Hirth.2024}, \cite{VaziriRad.2023}, \cite{VaziriRad.2024}\\
\midrule
Lack of cooling water for power plants & \gls{SBP} & \cite{Eisenack.2016}, \cite{Guenand.2024}, \cite{Jasiunas.2021}, \cite{Schmitz.2024}, \cite{Shinde.2023}, \cite{vanVliet.2016}, \cite{Wang.2022}, \cite{Wang.2023}\\
\midrule
War & \gls{SBP} & \cite{Banna.2023}, \cite{Bento.2024}, \cite{Braun.2023}, \cite{Braun.2024}, \cite{Luschini.2024}, \cite{McCollum.2020}, \cite{Nguyen.2024}\\
\bottomrule
\end{tabularx}
\end{center}
\end{table*}

\subsection{Impact of shock events and SBPs over time}
\label{subsec:impact_shock_SBP_over_time}

As discussed in \Cref{subsubsec:resilience}, a trapezoidal time curve often represents the resilience behaviour of electricity or energy systems. In such representations, time is plotted on the x-axis in these representations, while ``(energy) system performance'' is shown on the y-axis.
Typically, system performance declines following a \gls{HILP} event, remains at a reduced level for a specific duration, and subsequently recovers.
Some studies (e.g. \citealp{Braun.2020,Stankovic.2023}) suggest that post-recovery system performance may surpass pre-incident levels, as lessons learned from the event can enhance future resilience.
\par
\Cref{fig_shock_sbp_over_time} categorises system performance into three dimensions, which distinguishes it from the usual trapezoidal representations: technical, economic, and environmental.
The axes intentionally lack specific units and should be interpreted schematically.
The dotted horizontal line indicates system performance prior to a shock event (\Cref{fig_shock_sbp_over_time_a}, \Cref{fig_shock_sbp_over_time_c}, and \Cref{fig_shock_sbp_over_time_e}) or \gls{SBP} (\Cref{fig_shock_sbp_over_time_b}, \Cref{fig_shock_sbp_over_time_d}, and \Cref{fig_shock_sbp_over_time_f}).
The area between the initial state and the corresponding coloured curve represents the disturbance intensity (illustrated in \Cref{fig_shock_sbp_over_time_a}), which depends on both the duration (horizontal axis) and the degradation of system performance (vertical axis).
The latter, often referred to as severity \citep{CIGRE.2017}, reflects the magnitude of the performance loss.
For illustration, we consider an earthquake that damages a nuclear power plant as an example of a shock event.
As an example of an \gls{SBP}, we consider a gradual reduction in cooling water availability at the same facility, worsening over time.
\par
We generally assume that if a shock event or an \gls{SBP} is foreseeable, there will be time for preparation (between $t_1$ \mbox{and $t_2$)}. Such preparation can reduce the technical, economic, and environmental consequences for the energy system.
Accordingly, a timeline is presented that distinguishes between scenarios with preparatory measures (``w/ preparation'') and without preparatory measures (``w/o preparation'').
As illustrated by the blue and orange integrals in \Cref{fig_shock_sbp_over_time_b}, disturbance intensity is greater in cases without preparation.
\par
An \gls{SBP} may or may not trigger a subsequent shock event (between $t_2$ \mbox{and $t_3$)}.
For example, the aforementioned reduced power plant availability due to a lack of cooling water could eventually lead to a complete shutdown.
We further posit that a sudden shock event will generally cause greater disturbance intensity across all three dimensions compared to an \gls{SBP}.
If such a shock event occurs, performance degradation is most severe between $t_3$ \mbox{and $t_4$}, regardless of whether the shock event was preceded by an \gls{SBP}.
However, by $t_6$, the system is expected to return to its best possible state under all scenarios.

\newcommand{\seqShockSBP}{
\begin{tikzpicture}
\draw[->] (0,0) -- (8,0) node[below] {}; 
\draw[->] (0,0) -- (0,5) node[midway, above, rotate=90] {\yaxislabel}; 

\node[anchor=center] at (2.0, 0) {\includegraphics[height=0.3cm]{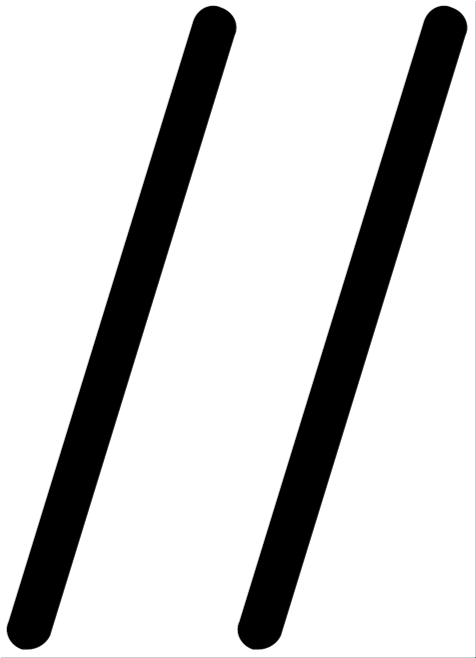}};

\definecolor{fhgreen}{RGB}{23,156,125}
\definecolor{fhblue}{RGB}{0,91,127}
\definecolor{fhgrey}{RGB}{166,187,200}
\definecolor{fhorange}{RGB}{245,130,32}
\definecolor{fhpurple}{RGB}{124,21,77}
\definecolor{fhmix}{RGB}{123,111,80}

\foreach \x in {1,3,3.4,3.9,5,7} {
    \draw[dotted, thin] (\x,0) -- (\x,5);
}

\foreach \y in {4} {
    \draw[dotted, thin] (0,\y) -- (8,\y);
}

\node at (1,-0.3) {$t_1$};
\node at (3,-0.3) {$t_2$};
\node at (3.4,-0.3) {$t_3$};
\node at (3.9,-0.3) {$t_4$};
\node at (5,-0.3) {$t_5$};
\node at (7,-0.3) {$t_6$};

\fill [fhgreen!25] \areacoordinatesone;
\node[align=center, text=black!50] at \areatextcoordinatesone {\areatextone};

\fill [fhblue!25] \areacoordinatestwo;
\node[align=center, text=black!50] at \areatextcoordinatestwo {\areatexttwo};
\fill [fhorange!25] \areacoordinatesthree;
\node[align=center, text=black!50] at \areatextcoordinatesthree {\areatextthree};
\fill [fhmix!25] \areacoordinatesfour;
\node[align=center, text=black!50] at \areatextcoordinatesfour {\areatextfour};

\fill [fhblue!25] \areacoordinatesfive;
\fill [fhorange!25] \areacoordinatessix;

\draw \colorwprep \wprep;
\draw \colorwoprep \woprep;

\draw \colorwprep \coordwpreplegend node[right] {w/ preparation};
\draw \colorwoprep \coordwopreplegend node[right] {w/o preparation};
\draw \colorwprep \areawpreplegend;
\draw \colorwoprep \areawopreplegend;

\node[align=center] at \textboxpositionone {\textboxone};
\draw [->] \drawarrowone;
\node[align=center] at \textboxpositiontwo {\textboxtwo};
\draw [->] \drawarrowtwo;
\node[align=center] at \textboxpositionthree {\textboxthree};
\draw [->] \drawarrowthree;
\node[align=center] at \textboxpositionfour {\textboxfour};

\end{tikzpicture}
}
\begin{figure*}[h]
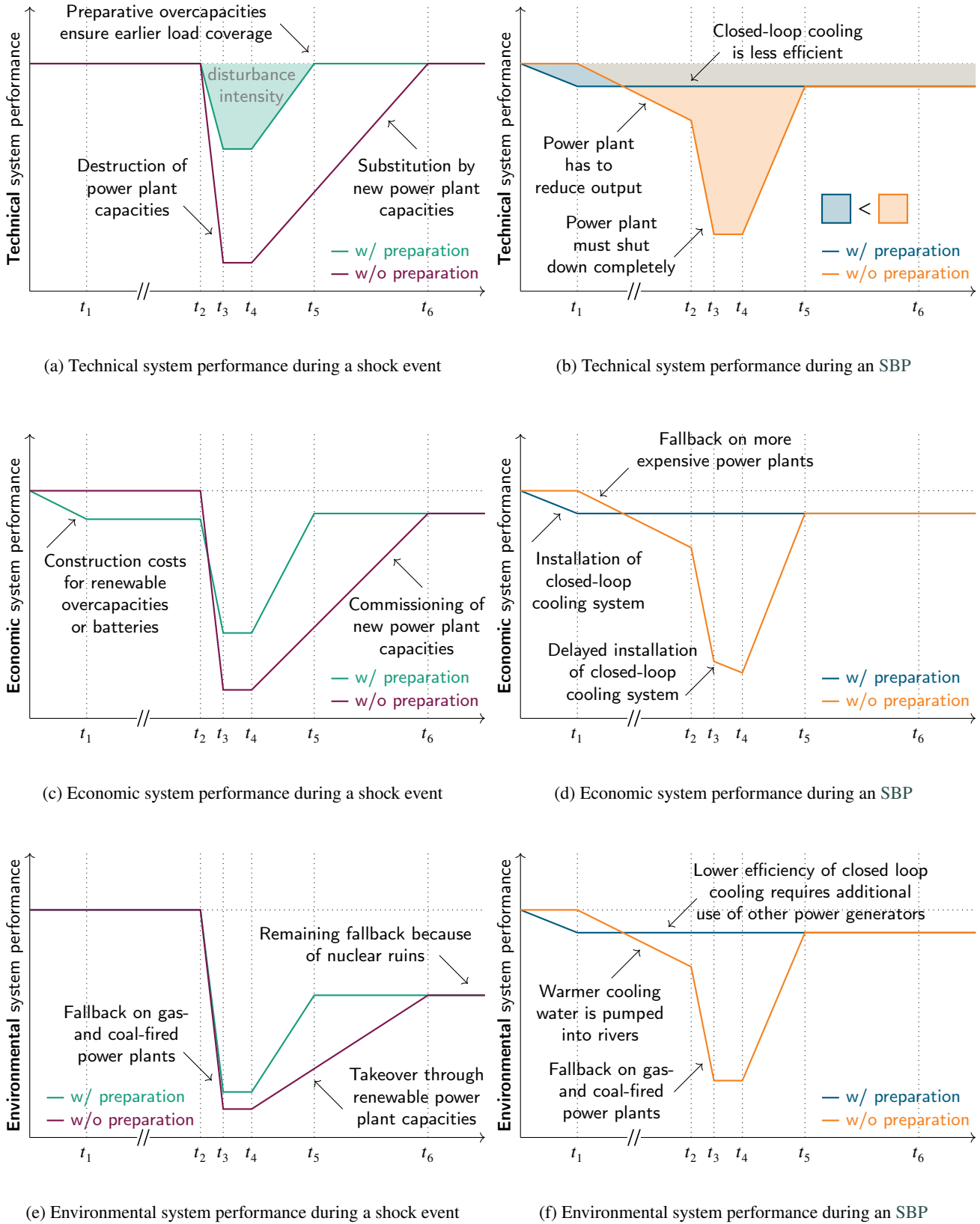

    \centering
    \begin{subfigure}[b]{0.49\textwidth}
         \centering
         \newcommand{\yaxislabel}{\textbf{Technical} system performance} 
         \newcommand{\colorwprep}{[thick,fhgreen]}
         \newcommand{\colorwoprep}{[thick,fhpurple]}
         \newcommand{\wprep}{(0,4) -- (1,4) -- (3,4) -- (3.4,2.5) -- (3.9,2.5) -- (5,4) -- (7,4) -- (8,4)} 
         \newcommand{\woprep}{(0,4) -- (1,4) -- (3,4) -- (3.4,0.5) -- (3.9,0.5) -- (7,4) -- (8,4)} 
         \newcommand{\coordwpreplegend}{(5.3,0.7) -- (5.6,0.7)}
         \newcommand{\coordwopreplegend}{(5.3,0.3) -- (5.6,0.3)}
         \newcommand{\textboxpositionone}{(1.8,1.8)}
         \newcommand{\textboxone}{Destruction of\\power plant\\capacities}
         \newcommand{\drawarrowone}{(2.7,1.5) -- (3.2,1.0)}
         \newcommand{\textboxpositiontwo}{(6.8,1.8)}
         \newcommand{\textboxtwo}{Substitution by\\new power plant\\capacities}
         \newcommand{\drawarrowtwo}{(6.8,2.5) -- (6.3,3.0)}
         \newcommand{\textboxpositionthree}{(2.4,4.7)}
         \newcommand{\textboxthree}{Preparative overcapacities\\ensure earlier load coverage}
         \newcommand{\drawarrowthree}{(4.4,4.7) -- (4.9,4.2)}
         \newcommand{\textboxpositionfour}{(0,0)} 
         \newcommand{\textboxfour}{} 
         \newcommand{\areacoordinatesone}{(3,4) -- (3.4,2.5) -- (3.9,2.5) -- (5,4) -- cycle} 
         \newcommand{\areatextcoordinatesone}{(3.9,3.6)} 
         \newcommand{\areatextone}{disturbance\\intensity} 
         \newcommand{\areacoordinatestwo}{(0,0)} 
         \newcommand{\areatextcoordinatestwo}{(0,0)} 
         \newcommand{\areatexttwo}{} 
         \newcommand{\areacoordinatesthree}{(0,0)} 
         \newcommand{\areatextcoordinatesthree}{(0,0)} 
         \newcommand{\areatextthree}{} 
         \newcommand{\areacoordinatesfour}{(0,0)} 
         \newcommand{\areatextcoordinatesfour}{(0,0)} 
         \newcommand{\areatextfour}{} 
         \newcommand{\areacoordinatesfive}{(0,0)} 
         \newcommand{\areacoordinatessix}{(0,0)} 
         \newcommand{\areawpreplegend}{(0,0)} 
         \newcommand{\areawopreplegend}{(0,0)} 
         \seqShockSBP{}
         \caption{Technical system performance during a shock event}
         \label{fig_shock_sbp_over_time_a}
    \end{subfigure}
    \vspace{0.8cm} 
    \begin{subfigure}[b]{0.49\textwidth}
         \centering
         \newcommand{\yaxislabel}{\textbf{Technical} system performance} 
         \newcommand{\colorwprep}{[thick,fhblue]}
         \newcommand{\colorwoprep}{[thick,fhorange]}
         \newcommand{\wprep}{(0,4) -- (1,3.6) -- (3,3.6) -- (8,3.6)} 
         \newcommand{\woprep}{(0,4) -- (1,4) -- (3,3) -- (3.4,1.0) -- (3.9,1.0) -- (5,3.6) -- (8,3.6)} 
         \newcommand{\coordwpreplegend}{(5.3,0.7) -- (5.6,0.7)}
         \newcommand{\coordwopreplegend}{(5.3,0.3) -- (5.6,0.3)}
         \newcommand{\textboxpositionone}{(1.2,2.2)}
         \newcommand{\textboxone}{Power plant\\has to\\reduce output}
         \newcommand{\drawarrowone}{(1.7,2.8) -- (2.2,3.3)}
         \newcommand{\textboxpositiontwo}{(1.6,0.8)}
         \newcommand{\textboxtwo}{Power plant\\must shut\\down completely}
         \newcommand{\drawarrowtwo}{(2.7,0.6) -- (3.2,1.1)}
         \newcommand{\textboxpositionthree}{(4.7,4.4)}
         \newcommand{\textboxthree}{Closed-loop cooling\\is less efficient}
         \newcommand{\drawarrowthree}{(3.5,4.2) -- (3.0,3.7)}
         \newcommand{\textboxpositionfour}{(6.05,1.45)} 
         \newcommand{\textboxfour}{<} 
         \newcommand{\areacoordinatesone}{(0,0)} 
         \newcommand{\areatextcoordinatesone}{(0,0)} 
         \newcommand{\areatextone}{} 
         \newcommand{\areacoordinatestwo}{(0,4) -- (1,3.6) -- (1.8,3.6) -- (1,4) -- cycle} 
         \newcommand{\areatextcoordinatestwo}{(0,0)} 
         \newcommand{\areatexttwo}{} 
         \newcommand{\areacoordinatesthree}{(1.8,3.6) -- (3,3) -- (3.4,1.0) -- (3.9,1.0) -- (5,3.6) -- cycle} 
         \newcommand{\areatextcoordinatesthree}{(0,0)} 
         \newcommand{\areatextthree}{} 
         \newcommand{\areacoordinatesfour}{(1,4) -- (1.8,3.6) -- (8,3.6) -- (8,4) -- cycle} 
         \newcommand{\areatextcoordinatesfour}{(0,0)} 
         \newcommand{\areatextfour}{} 
         \newcommand{\areacoordinatesfive}{(5.3,1.2) -- (5.3,1.7) -- (5.8,1.7) -- (5.8,1.2) -- cycle}
         \newcommand{\areacoordinatessix}{(6.3,1.2) -- (6.3,1.7) -- (6.8,1.7) -- (6.8,1.2) -- cycle}
         \newcommand{\areawpreplegend}{(5.3,1.2) -- (5.3,1.7) -- (5.8,1.7) -- (5.8,1.2) -- cycle}
         \newcommand{\areawopreplegend}{(6.3,1.2) -- (6.3,1.7) -- (6.8,1.7) -- (6.8,1.2) -- cycle}
         \seqShockSBP{}
         \caption{Technical system performance during an \gls{SBP}}
         \label{fig_shock_sbp_over_time_b}
    \end{subfigure}
    \hfill
    \begin{subfigure}[b]{0.49\textwidth}
         \centering
         \newcommand{\yaxislabel}{\textbf{Economic} system performance} 
         \newcommand{\colorwprep}{[thick,fhgreen]}
         \newcommand{\colorwoprep}{[thick,fhpurple]}
         \newcommand{\wprep}{(0,4) -- (1,3.5) -- (3,3.5) -- (3.4,1.5) -- (3.9,1.5) -- (5,3.6) -- (8,3.6)} 
         \newcommand{\woprep}{(0,4) -- (1,4) -- (3,4) -- (3.4,0.5) -- (3.9,0.5) -- (7,3.6) -- (8,3.6)} 
         \newcommand{\coordwpreplegend}{(5.3,0.7) -- (5.6,0.7)}
         \newcommand{\coordwopreplegend}{(5.3,0.3) -- (5.6,0.3)}
         \newcommand{\textboxpositionone}{(1.5,2.2)}
         \newcommand{\textboxone}{Construction costs\\for renewable\\overcapacities\\or batteries}
         \newcommand{\drawarrowone}{(0.3,3.0) -- (0.8,3.5)}
         \newcommand{\textboxpositiontwo}{(6.8,1.6)}
         \newcommand{\textboxtwo}{Commissioning of\\new power plant\\capacities}
         \newcommand{\drawarrowtwo}{(6.9,2.3) -- (6.4,2.8)}
         \newcommand{\textboxpositionthree}{(0,0)} 
         \newcommand{\textboxthree}{} 
         \newcommand{\drawarrowthree}{} 
         \newcommand{\textboxpositionfour}{(0,0)} 
         \newcommand{\textboxfour}{} 
         \newcommand{\areacoordinatesone}{(0,0)} 
         \newcommand{\areatextcoordinatesone}{(0,0)} 
         \newcommand{\areatextone}{} 
         \newcommand{\areacoordinatestwo}{(0,0)} 
         \newcommand{\areatextcoordinatestwo}{(0,0)} 
         \newcommand{\areatexttwo}{} 
         \newcommand{\areacoordinatesthree}{(0,0)} 
         \newcommand{\areatextcoordinatesthree}{(0,0)} 
         \newcommand{\areatextthree}{} 
         \newcommand{\areacoordinatesfour}{(0,0)} 
         \newcommand{\areatextcoordinatesfour}{(0,0)} 
         \newcommand{\areatextfour}{} 
         \newcommand{\areacoordinatesfive}{(0,0)} 
         \newcommand{\areacoordinatessix}{(0,0)} 
         \newcommand{\areawpreplegend}{(0,0)} 
         \newcommand{\areawopreplegend}{(0,0)} 
         \seqShockSBP{}
         \caption{Economic system performance during a shock event}
         \label{fig_shock_sbp_over_time_c}
    \end{subfigure}
    \vspace{0.8cm} 
    \begin{subfigure}[b]{0.49\textwidth}
         \centering
         \newcommand{\yaxislabel}{\textbf{Economic} system performance} 
         \newcommand{\colorwprep}{[thick,fhblue]}
         \newcommand{\colorwoprep}{[thick,fhorange]}
         \newcommand{\wprep}{(0,4) -- (1,3.6) -- (3,3.6) -- (8,3.6)} 
         \newcommand{\woprep}{(0,4) -- (1,4) -- (3,3) -- (3.4,1.0) -- (3.9,0.8) -- (5,3.6) -- (8,3.6)} 
         \newcommand{\coordwpreplegend}{(5.3,0.7) -- (5.6,0.7)}
         \newcommand{\coordwopreplegend}{(5.3,0.3) -- (5.6,0.3)}
         \newcommand{\textboxpositionone}{(1.2,2.4)}
         \newcommand{\textboxone}{Installation of\\closed-loop\\cooling system}
         \newcommand{\drawarrowone}{(0.2,3.1) -- (0.7,3.6)}
         \newcommand{\textboxpositiontwo}{(3.6,4.7)}
         \newcommand{\textboxtwo}{Fallback on more\\expensive power plants}
         \newcommand{\drawarrowtwo}{(1.9,4.4) -- (1.4,3.9)}
         \newcommand{\textboxpositionthree}{(1.8,0.8)}
         \newcommand{\textboxthree}{Delayed installation\\of closed-loop\\cooling system}
         \newcommand{\drawarrowthree}{(2.9,0.4) -- (3.4,0.9)}
         \newcommand{\textboxpositionfour}{(0,0)} 
         \newcommand{\textboxfour}{} 
         \newcommand{\areacoordinatesone}{(0,0)} 
         \newcommand{\areatextcoordinatesone}{(0,0)} 
         \newcommand{\areatextone}{} 
         \newcommand{\areacoordinatestwo}{(0,0)} 
         \newcommand{\areatextcoordinatestwo}{(0,0)} 
         \newcommand{\areatexttwo}{} 
         \newcommand{\areacoordinatesthree}{(0,0)} 
         \newcommand{\areatextcoordinatesthree}{(0,0)} 
         \newcommand{\areatextthree}{} 
         \newcommand{\areacoordinatesfour}{(0,0)} 
         \newcommand{\areatextcoordinatesfour}{(0,0)} 
         \newcommand{\areatextfour}{} 
         \newcommand{\areacoordinatesfive}{(0,0)} 
         \newcommand{\areacoordinatessix}{(0,0)} 
         \newcommand{\areawpreplegend}{(0,0)} 
         \newcommand{\areawopreplegend}{(0,0)} 
         \seqShockSBP{}
         \caption{Economic system performance during an \gls{SBP}}
         \label{fig_shock_sbp_over_time_d}
    \end{subfigure}
    \hfill
    \begin{subfigure}[b]{0.49\textwidth}
         \centering
         \newcommand{\yaxislabel}{\textbf{Environmental} system performance} 
         \newcommand{\colorwprep}{[thick,fhgreen]}
         \newcommand{\colorwoprep}{[thick,fhpurple]}
         \newcommand{\wprep}{(0,4) -- (1,4) -- (3,4) -- (3.4,0.8) -- (3.9,0.8) -- (5,2.5) -- (8,2.5)} 
         \newcommand{\woprep}{(0,4) -- (1,4) -- (3,4) -- (3.4,0.5) -- (3.9,0.5) -- (7,2.5) -- (8,2.5)} 
         \newcommand{\coordwpreplegend}{(0.3,0.7) -- (0.6,0.7)}
         \newcommand{\coordwopreplegend}{(0.3,0.3) -- (0.6,0.3)}
         \newcommand{\textboxpositionone}{(1.7,1.8)}
         \newcommand{\textboxone}{Fallback on gas-\\and coal-fired\\power plants}
         \newcommand{\drawarrowone}{(2.7,1.5) -- (3.2,1.0)}
         \newcommand{\textboxpositiontwo}{(6.8,0.7)}
         \newcommand{\textboxtwo}{Takeover through\\renewable power\\plant capacities}
         \newcommand{\drawarrowtwo}{(5.5,0.5) -- (5.0,1.0)}
         \newcommand{\textboxpositionthree}{(5.9,3.4)}
         \newcommand{\textboxthree}{Remaining fallback because\\of nuclear ruins}
         \newcommand{\drawarrowthree}{(7.2,3.2) -- (7.7,2.7)}
         \newcommand{\textboxpositionfour}{(0,0)} 
         \newcommand{\textboxfour}{} 
         \newcommand{\areacoordinatesone}{(0,0)} 
         \newcommand{\areatextcoordinatesone}{(0,0)} 
         \newcommand{\areatextone}{} 
         \newcommand{\areacoordinatestwo}{(0,0)} 
         \newcommand{\areatextcoordinatestwo}{(0,0)} 
         \newcommand{\areatexttwo}{} 
         \newcommand{\areacoordinatesthree}{(0,0)} 
         \newcommand{\areatextcoordinatesthree}{(0,0)} 
         \newcommand{\areatextthree}{} 
         \newcommand{\areacoordinatesfour}{(0,0)} 
         \newcommand{\areatextcoordinatesfour}{(0,0)} 
         \newcommand{\areatextfour}{} 
         \newcommand{\areacoordinatesfive}{(0,0)} 
         \newcommand{\areacoordinatessix}{(0,0)} 
         \newcommand{\areawpreplegend}{(0,0)} 
         \newcommand{\areawopreplegend}{(0,0)} 
         \seqShockSBP{}
         \caption{Environmental system performance during a shock event}
         \label{fig_shock_sbp_over_time_e}
    \end{subfigure}
    \vspace{0.6cm} 
    \begin{subfigure}[b]{0.49\textwidth}
         \centering
         \newcommand{\yaxislabel}{\textbf{Environmental} system performance} 
         \newcommand{\colorwprep}{[thick,fhblue]}
         \newcommand{\colorwoprep}{[thick,fhorange]}
         \newcommand{\wprep}{(0,4) -- (1,3.6) -- (3,3.6) -- (8,3.6)} 
         \newcommand{\woprep}{(0,4) -- (1,4) -- (3,3) -- (3.4,1.0) -- (3.9,1.0) -- (5,3.6) -- (8,3.6)} 
         \newcommand{\coordwpreplegend}{(5.3,0.7) -- (5.6,0.7)}
         \newcommand{\coordwopreplegend}{(5.3,0.3) -- (5.6,0.3)}
         \newcommand{\textboxpositionone}{(1.6,0.8)}
         \newcommand{\textboxone}{Fallback on gas-\\and coal-fired\\power plants}
         \newcommand{\drawarrowone}{(2.7,0.8) -- (3.2,1.3)}
         \newcommand{\textboxpositiontwo}{(1.4,2.2)}
         \newcommand{\textboxtwo}{Warmer cooling\\water is pumped\\into rivers}
         \newcommand{\drawarrowtwo}{(1.6,2.8) -- (2.1,3.3)}
         \newcommand{\textboxpositionthree}{(5.1,4.3)}
         \newcommand{\textboxthree}{Lower efficiency of closed loop\\cooling requires additional\\use of other power generators}
         \newcommand{\drawarrowthree}{(3.1,4.2) -- (2.6,3.7)}
         \newcommand{\textboxpositionfour}{(0,0)} 
         \newcommand{\textboxfour}{} 
         \newcommand{\areacoordinatesone}{(0,0)} 
         \newcommand{\areatextcoordinatesone}{(0,0)} 
         \newcommand{\areatextone}{} 
         \newcommand{\areacoordinatestwo}{(0,0)} 
         \newcommand{\areatextcoordinatestwo}{(0,0)} 
         \newcommand{\areatexttwo}{} 
         \newcommand{\areacoordinatesthree}{(0,0)} 
         \newcommand{\areatextcoordinatesthree}{(0,0)} 
         \newcommand{\areatextthree}{} 
         \newcommand{\areacoordinatesfour}{(0,0)} 
         \newcommand{\areatextcoordinatesfour}{(0,0)} 
         \newcommand{\areatextfour}{} 
         \newcommand{\areacoordinatesfive}{(0,0)} 
         \newcommand{\areacoordinatessix}{(0,0)} 
         \newcommand{\areawpreplegend}{(0,0)} 
         \newcommand{\areawopreplegend}{(0,0)} 
         \seqShockSBP{}
         \caption{Environmental system performance during an \gls{SBP}}
         \label{fig_shock_sbp_over_time_f}
    \end{subfigure}
    \caption{Technical, economic, and environmental system performances during a shock event (left side, using the example of an earthquake that destroys a nuclear power plant) and during an \gls{SBP} (right side, using the example of a lack of cooling water in a nuclear power plant that becomes more severe over time), own illustration.}
    \label{fig_shock_sbp_over_time}
    \vspace{-1.0em}
\end{figure*}

It is important to note that the horizontal axis in these examples includes breaks and is not to scale.
While \glspl{SBP} may develop over several years, shock events can unfold within seconds or minutes.
Additionally, the illustrated sequences serve as examples and do not apply to all possible events.

\subsubsection{Example sequence of a shock event}
\Cref{fig_shock_sbp_over_time_a} illustrates the technical performance dynamics of an energy system experiencing a shock event caused by an earthquake destroying a nuclear power plant.
Without preparatory measures, the technical performance undergoes a sudden and severe decline.
The reconstruction of new power plant capacities takes time, resulting in a prolonged period of diminished performance until sufficient capacity is gradually commissioned, ultimately restoring the system's ability to meet total electricity demand.
Since earthquake prediction is not yet possible \citep{Akhoondzadeh.2024}, the only viable preparatory measure is the construction of overcapacities.
Although this does not improve the technical performance before the event, as the additional electricity cannot be consumed, it mitigates the overall intensity of disruption following the nuclear power plant's destruction.
As a result, system recovery occurs at $t_5$ instead of $t_6$.
\par
As shown in \Cref{fig_shock_sbp_over_time_c}, the economic performance of the system experiences a sharp decline immediately after the shock event if no preparatory measures are taken ($t_2$ to $t_3$).
The power plant incurs immediate depreciation coupled with potential costs from unmet electricity demands.
This economic state persists ($t_3$ to $t_4$) until performance begins to improve with the phased commissioning of new power plant capacities ($t_4$ to $t_6$).
In contrast, the construction of renewable over-capacities initially deteriorates economic performance, incurring upfront costs without immediate demand.
However, during the shock event this measure results in a less severe economic decline and a faster recovery as the nuclear power plant must still be fully depreciated.
The burden of uncovered loads is reduced.
In both cases, the economic performance fails to return to pre-shock levels primarily due to the long-term costs associated with dismantling nuclear ruins.
\par
The system's environmental performance (\Cref{fig_shock_sbp_over_time_e}) also experiences a sharp decline in the absence of preparatory measures, as environmental contamination occurs and as we assume that CO\textsubscript{2}-emitting gas- and coal-fired power plants must compensate for the lost capacity.
It is assumed that subsequent over-capacity investments will be renewable, leading to a gradual improvement in environmental performance over time.
The preparatory construction of over-capacities helps mitigate the environmental decline, as fewer alternative power plants are needed.
Nonetheless, similar to economic performance, the environmental performance does not fully recover to its initial state, due to the long-term ecological damage from the nuclear power plant's destruction.

\subsubsection{Example sequence of an \gls{SBP}}
In the \gls{SBP} example involving a gradual reduction of cooling water availability in a nuclear power plant, the decline in technical performance (illustrated in \Cref{fig_shock_sbp_over_time_b}) begins earlier, at time $t_1$, compared to the earthquake-induced shock event, which occurs at $t_2$.
Without preparatory measures, the technical performance steadily decreases as cooling water becomes increasingly scarce due to rising ambient temperatures.
This decline culminates in a final shock event between $t_2$ \mbox{and $t_3$}, where insufficient cooling water forces a complete power plant shutdown.
Installing a closed-loop cooling system as a preparatory measure improves technical performance but does not fully restore it to pre-\gls{SBP} levels, as closed-loop cooling systems are less efficient and reduce the power plant's available capacity \citep{Byers.2014}.
Nonetheless, early implementation of such a system mitigates the severity of technical decline and prevents a total shutdown.
\par
In the absence of preparatory measures, the system's economic performance (\Cref{fig_shock_sbp_over_time_d}) initially mirrors the technical performance.
The gradual decline in available capacity, eventually leading to complete failure, necessitates reliance on costlier replacement power plants.
Subsequently, additional expenses arise from the late installation of a closed-loop cooling system.
Conversely, proactively installing the closed-loop system incurs upfront costs but avoids the higher financial burdens associated with a full outage.
\par
In line with previous trends, environmental system performance (\Cref{fig_shock_sbp_over_time_f}) also deteriorates due to the shock event, as warmer cooling water is discharged from the power plant into the river.
Moreover, relying on gas- and coal-fired replacement power plants causes higher CO\textsubscript{2} emissions.
The lower efficiency of the closed-loop cooling system also requires supplementary power plants after installation, the construction and operation of which have additional environmental consequences.
However, the closed-loop cooling system reduces the volume of warmer cooling water discharged into the river, thereby improving environmental performance.
\par
When comparing the \gls{SBP} without preparatory measures to the earthquake-induced shock event, the declines in technical, economic, and environmental performance are less severe, primarily because the power plant remains intact.
Consequently, the system recovers more quickly, resolving the aftermath by $t_5$ rather than $t_6$.
\section{Resilience perspectives for integrated energy system planning}
\label{sec:transformation_pathway}

A future challenge facing energy system planners is the application of resilience capacities to transformation pathways of complex integrated energy systems.
A central consideration is identifying the appropriate levels and dimensions for enhancing system resilience, as well as determining how traditional energy system models can support the identification of economically viable and secure solutions.
\par
To broaden and advance planning perspectives, this section explores how resilience capacities can be embedded into planning frameworks for integrated energy systems.
\Cref{subsec:resilience_capacities} presents the fundamental resilience capacities, followed by detailed explanations of the identified recourse options: absorptive (\Cref{subsec:absorptive_recourse_options}), adaptive (\Cref{subsec:adaptive_recourse_options}), and transformative (\Cref{subsec:transformative_recourse_options}) measures.
We further discuss which markets encourage investments in the respective recourse options.


\subsection{Resilience capacities}
\label{subsec:resilience_capacities}


The ``shocks and capacities'' concept from \cite{Manca.2017} provides an overview of different resilience capacities.
Based on the disturbance intensity and its duration of exposure, the authors classify these capacities into three levels:

\begin{enumerate}
    \item \textbf{Absorptive capacity} (first level) -- the ability to withstand minor disturbances with normal operational system adjustments.
    \item \textbf{Adaptive capacity} (second level) -- the ability to adjust operations to accommodate more significant and/or extended disturbances.
    \item \textbf{Transformative capacity} (third level) -- the ability to fundamentally change system configurations to manage upcoming extreme and/or prolonged disturbances.
\end{enumerate}

\begin{figure}[ht]
    \centering
    \begin{tikzpicture}

\definecolor{fhgreen}{RGB}{23,156,125}
\definecolor{fhblue}{RGB}{0,91,127}
\definecolor{fhgrey}{RGB}{166,187,200}
\definecolor{fhorange}{RGB}{245,130,32}
\definecolor{fhpurple}{RGB}{124,21,77}
\definecolor{lightgrey}{RGB}{196,210,218}


\setstretch{0.8} 

\shade [bottom color=fhgreen, top color=fhgrey, shading angle=-45, opacity=.2] (5.8,0,0) arc (0:90:5.8) -- (0,0,0) -- cycle;
\shade [bottom color=fhblue, top color=fhgreen, shading angle=-45, opacity=.2] (3.5,0,0) arc (0:90:3.5) -- (0,0,0) -- cycle;

\draw[fhblue] (0,0) -- (0:5.8) arc (0:90:5.8) -- cycle;
\draw[fhgreen] (0,0) -- (0:3.5) arc (0:90:3.5) -- cycle;

\draw[->] (0,0) -- (7,0); 
\draw[->] (0,0) -- (0,7); 

\node[align=center] at (3.495,-0.7) {\textcolor{black}{\textbf{Planning horizon}}};

\node[align=center] at (1.75,-0.25) {\textcolor{black}{Operational}};
\node[align=center] at (4.65,-0.25) {\textcolor{black}{Curative}};
\node[align=center] at (6.3,-0.25) {\textcolor{black} {Investment}};

\node[align=center,rotate=90] at (-0.70,3.495) {\textcolor{black}{\textbf{Resilience capacity}}};
\node[align=center,rotate=90] at (-0.25,1.75) {\textcolor{black}{Absorb}};
\node[align=center,rotate=90] at (-0.25,4.65) {\textcolor{black}{Adapt}};
\node[align=center,rotate=90] at (-0.25,6.3) {\textcolor{black}{Transform}};

\node[align=center] at (1.6,0.4) {\scriptsize \textbf{Curtailing}\\\scriptsize \acrshort{VRE} production, \acrshort{XB} flows};


\node[align=center] at (0.8,1.4) {\scriptsize \textbf{Adjusting}\\\scriptsize thermal\\\scriptsize generation};

\node[align=center] at (2.3,1.3) {\scriptsize \textbf{Transmission}\\\scriptsize \textbf{switching}};

\node[align=center] at (1.0,2.6) {\scriptsize \textbf{Shifting}\\\scriptsize loads in various\\\scriptsize end-use sectors};

\node[align=center] at (1.25,4.8) {\scriptsize \textbf{Shedding/rationing}\\\scriptsize loads in various \\\scriptsize end-use sectors};

\node[align=center] at (4.4,1.5) {\scriptsize \textbf{Utilising}\\\scriptsize hybrid or multivalent\\\scriptsize supply systems};

\node[align=center] at (3.8,2.7) {\scriptsize \textbf{Importing more}\\\scriptsize combustible/feedstock \\\scriptsize fuels, gas, electricity};

\node[align=center] at (2.8,3.8) {\scriptsize \textbf{Activating}\\\scriptsize long-term storage};

\node[align=center] at (4.6,0.4) {\scriptsize \textbf{Utilising}\\\scriptsize firm capacities};


\node[align=center] at (1.5,6.3) {\scriptsize \textbf{Spatially (re-)allocating}\\\scriptsize (de-)centralised\\\scriptsize supply/demand};

\node[align=center] at (3.5,5.8) {\scriptsize \textbf{Developing}\\\scriptsize new technologies\\\scriptsize and combinations};

\node[align=center] at (4.9,4.9) {\scriptsize \textbf{Hybridising}\\\scriptsize supply systems};

\node[align=center] at (5.7,4.1) {\scriptsize \textbf{Deploying}\\\scriptsize storage technologies};

\node[align=center] at (6.1,3.2) {\scriptsize \textbf{Scaling up}\\\scriptsize power generation\\\scriptsize capacities};

\node[align=center] at (6.3,2.1) {\scriptsize \textbf{Expanding}\\\scriptsize energy\\\scriptsize transport};

\node[align=center] at (6.5,0.8) {\scriptsize \textbf{Adopting}\\\scriptsize negative\\\scriptsize emission\\\scriptsize technology};

\node[align=center] at (5.8,5.8) {\textbf{...}};

\end{tikzpicture}
    \caption{Resilience capacities of integrated energy systems in response to shock events and \glspl{SBP}, own illustration based on the ``shocks and capacities'' framework by \cite{Manca.2017}. Note that the options across all levels are illustrative, not exhaustive.}
    \label{fig_capacity_concept}
    \vspace{-1.0em}
\end{figure}
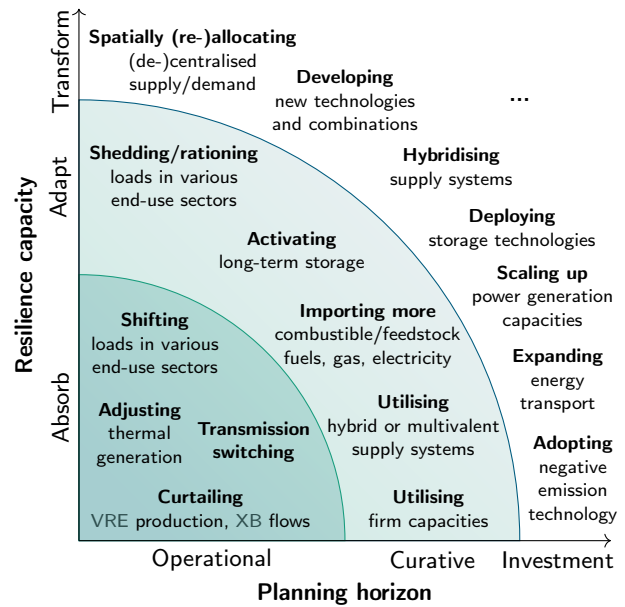

Drawing inspiration from this conceptual framework, \Cref{fig_capacity_concept} compiles various recourse options aimed at enhancing resilience in the context of integrated energy system planning.
This extension serves as guidance, broadening perspectives in system planning tasks by explicitly incorporating resilience capacities and associated recourse measures.
More specifically, it qualitatively represents the abilities of an increasingly integrated energy system (or model) to respond to previously defined shock events and \glspl{SBP} discussed in \Cref{sec:shocks_SBPs}.

The qualitative categorisation translates the time of exposure into different system planning horizons ranging from short-term operational, medium-term curative, to long-term investment planning measures.
In each category, we arrange the measures according to their time horizon and disturbance intensity, with decisions placed further outward in the diagram indicating greater time requirements and/or higher implementation efforts.

\subsection{Absorptive recourse options}
\label{subsec:absorptive_recourse_options}

The first level emphasises the absorptive capacity of integrated energy systems, aiming to maintain persistence with relatively minimal or standard effort \citep{Mentges.2023}.
It comprises recourse options designed to address minor, short-duration disturbances.
Examples include:
\begin{itemize}
    \item \textbf{Curtailment}: Temporarily reducing output from \gls{VRE} sources or limiting \gls{XB} trading capacity in response to excess electricity supply or alleviate transmission line congestion.
    \item \textbf{Transmission switching}: Controlling electricity flows within the transmission network by selectively opening or closing transmission assets.
    \item \textbf{Adjusting thermal generation}: Utilising peaking units and operating combined heat-and-power plants in condensing mode, or running thermal generation plants at a generally less efficient operating point.
    \item \textbf{Load shifting}: Incorporating demand response to encourage consumers to adjust (advance/delay) their energy use in response to price signals, demand side management to actively coordinate spatio-temporal energy demand, temporal load shifting via short-term storages, and spatial load shifting (e.g. by data centre computing loads \cite{Riepin.2025}).
\end{itemize}
In the \gls{EU}, these absorptive capacities typically generate revenue through reserve markets and are controlled by the transmission or distribution grid operators \citep{Schittekatte.2020}.
Future market designs should ensure adequate incentives for maintaining sufficient reserve capacities.
In energy system planning --- which commonly employs hourly or longer temporal resolutions --- these short-term measures may partly play a minor role.
Consequently, rapid system restoration capacities and grid-forming inverter capacities essential for grid stability are not depicted in \Cref{fig_capacity_concept}, despite their criticality to overall system functionality.

\subsection{Adaptive recourse options}
\label{subsec:adaptive_recourse_options}

The second level features the adaptive capacity of integrated energy systems and includes the following response and recourse options:
\begin{itemize}
    \item \textbf{Load shedding or rationing}: Measures of last resort -- whether immediate or gradual, voluntary or enforced -- to stabilise the energy system or maintain partial system performance. These involve either disconnecting consumers entirely from energy infrastructure (e.g. electricity grids, heating networks, or gas systems) or limiting their supply during specific time windows or in affected regions.
    \item \textbf{Activating long-term storage}: Utilising energy reserves from hydrogen and gas storage systems, thermal storage solutions, and hydropower reservoirs to ensure system stability during extended disruptions.
    \item \textbf{Importing additional fuels and electricity}: Increasing the supply of electricity, gas, and other fuels by leveraging existing import infrastructure to address shortfalls in domestic energy production or availability.
    \item \textbf{Utilising hybrid supply}: Adjusting operational modes in hybrid or multivalent systems, e.g. hybrid heat pumps or hybrid boilers, to meet end-use demands by prioritising commodities with lower criticality for the system.
    \item \textbf{Utilising firm capacities}: Providing resources that can consistently deliver electricity as they are crucial for ensuring grid stability and meeting demand. This may also include deferring maintenance, even when it incurs more costs, or deploying back-up generation units.
\end{itemize}
These adaptive capacities typically generate revenue through day-ahead or intraday markets \citep{Schittekatte.2020}.
However, during significant disruptions, price signals in these markets can become distorted, e.g. through government-imposed price caps, reducing the incentive to invest in these stability measures \citep{Rüdinger202359}.

\subsection{Transformative recourse options}
\label{subsec:transformative_recourse_options}
The third and outermost level constitutes the transformative capacity, requiring recourse options that fundamentally alter the energy system's structure when its current state can no longer adequately respond to significant and long-lasting disturbances:
\begin{itemize}
    \item \textbf{Spatial (re-)allocation}: Centralising or decentralising supply sources and demands, transforming the spatial distribution of producers and consumers.
    Examples include the construction of new \gls{LNG} terminals \citep{Gritz.2024,Wiertz.2023} or the expansion of domestic electrolysers instead of increased hydrogen imports \citep{Frischmuth.2022}.
    \item \textbf{Developing new technologies and combinations}: Investment in research and development for innovative technology combinations and technology breakthroughs, e.g. nuclear fusion \citep{Tokimatsu.2003}, despite inherent uncertainties.
    \item \textbf{Hybridising supply systems}: Designing and deploying hybrid, multivalent supply systems with built-in redundancy to enable fuel-switching flexibility, such as combining a direct resistive heating element with a fuel-based boiler.
    \item \textbf{Deploying storage}: Expanding short- to long-term storage options for different commodities to enhance flexibility and reliability.
    \item \textbf{Scaling generation capacity}: Commissioning additional capacities for thermal and renewable generation (e.g. onshore and offshore wind, solar photovoltaics, hydrogen power plants), which may also include building fossil power plants as back-up.
    \item \textbf{Expanding energy transport}: Building and upgrading energy transport infrastructure, e.g. high-voltage alternating current and high-voltage direct current lines, pipelines, and shipping port capacities.
    \item \textbf{Adopting negative emission technology}: Implementing engineered solutions, e.g. direct air capture and bioenergy with carbon capture and storage, alongside natural climate solutions such as reforestation and wetland restoration.
\end{itemize}
In the existing European market design, transformative capacities generate revenue on day-ahead or intraday markets, or by participating in long-term markets \citep{Schittekatte.2020}.
However, due to the uncertainty associated with \gls{HILP} events, market signals might not provide sufficient incentives to guarantee resilience.
To that end, supplementary mechanisms such as reserve or capacity markets can play a crucial role in encouraging the necessary investments for transformative resilience measures.
\section{Implications for integrated energy system planning}
\label{sec:modelling}

The previous sections demonstrated the importance of both operational planning and investment planning in maintaining energy security and creating resilient energy systems.
\Cref{subsec:policy_implications} gives an overview of planning approaches that are currently used in energy system modelling.
\Cref{subsec:incorporating_shocks_SBPs_into_energy_system_transformations} shows how shock events and \glspl{SBP} can be incorporated into energy system transformations.
Then, \Cref{subsec:challenges_in_the_assessment_of_resilience} briefly outlines key challenges in the assessment of resilience in integrated energy systems.
Finally, \Cref{sec:trade_off} discusses the inherent trade-off between energy system costs and resilience and the need to further substantiate and quantify the underlying effects.

\subsection{System planning approaches}
\label{subsec:policy_implications}

In operational planning, critical aspects involve effectively managing short-term uncertainties, identify vulnerabilities during real-time development, and what recourse options are available to system operators \citep{Braun.2024,Moretti.2020}.
Typical vulnerabilities include threats to physical infrastructure of power and energy systems, threats targeting digital and communication layers, and systemic weaknesses resulting from interdependence failures, coordination challenges, or human errors.
Investment and expansion planning, particularly for multi-period transformation pathways, requires comprehensive approaches capable of assessing a broader range of possibilities over more extended time frames compared to operational planning.

\subsubsection{Forward-looking investment planning}

Forward-looking investment planning approaches for energy markets and infrastructures enhance the ability to anticipate and mitigate future risks.
Conventional deterministic energy system planning, which evaluates only one scenario at a time, limits its ability to address extraordinary events~\citep{McCollum.2020}.
To overcome this limitation, these planning approaches typically analyse a small set of scenarios independently~\citep{Braun.2024}.
However, a significant challenge remains in advancing energy system modelling to incorporate essential aspects of resilience into these approaches.
\par
Integrating uncertainty explicitly into decision-support processes allows system planners to better recognise and mitigate risks associated with decisions made under uncertainty~\citep{Hartel.2021}.
While the hedging decisions might not resemble any optimal decisions obtained for individual scenarios~\citep{Wallace.2000}, common features among deterministic planning approaches based on perfect information do not necessarily provide robust or least-regret recommendations for planners and policymakers.
\par
Forward-looking investment planning is already a crucial feature of sound system planning approaches.
With long lead times and lifespans, lumpy and capital-intensive structures, and the potential irreversibility of decisions that might result in stranded assets, energy infrastructure decisions must be made with an expectation of uncertain developments, ensuring the resilience and security of future systems.

\subsubsection{Resilient system development planning}

Planning resilient energy systems involves navigating complex system interactions and uncertainties, extending beyond merely capturing variability in generation and load \citep{Braun.2024}.
Looking at a myriad of potential shock events and \glspl{SBP}, forward-looking system development planning must incorporate these additional uncertainties and risks to identify the most effective strategies for integrating resilience into the system design \citep{UNECE.2022}.
\par
Established system planning processes, such as the ``Ten-Year Network Development Plan'' by \gls{ENTSO-E} in Europe or the ``Transmission Needs Study'' in the United States, are key policy instruments facilitating the planning process of developing energy systems and infrastructures.
However, these decision-making support frameworks must evolve to address energy systems’ growing complexity and uncertainty, embedding resilience and security explicitly into long-term planning strategies.
\par
For instance, policy measures such as deploying liquefied natural gas terminals in response to the gas crisis in Germany in 2022 \citep{Gritz.2024,Wiertz.2023} often lacked efficiency and foresight.
These inefficiencies and gaps in foresight are partly a result of swift decision-making required to address disruptive events, which prioritises short-term solutions over long-term planning.
Moreover, inadequate coordination and collaboration can cause individual response measures to inadvertently create unintended or counterproductive impacts on interconnected systems~\citep{Draghi.2024}.
\par
Hence, pressing policy questions emerge regarding the balance between short-term (reactive) goals such as strengthening energy security and maintaining long-term (proactive) goals such as the clean energy transition \citep{Kim.2025}.
Addressing this dual challenge highlights the necessity of proactive and forward-looking approaches, integrating energy security and resilience into the broader goals of clean energy transformations.
Building and maintaining absorptive, adaptive, and transformative capacities is thus essential for resilient energy system transformations. 
Expanding current planning frameworks to include manifold recourse options fosters new resilience paradigms in integrated energy system design.

\subsection{Incorporating shock events and \glspl{SBP} into energy system transformations}
\label{subsec:incorporating_shocks_SBPs_into_energy_system_transformations}

As discussed in \Cref{sec:shocks_SBPs}, shock events and \glspl{SBP} can challenge resilience in transformation and system development pathways.
\Cref{subsubsec:transformation_pathways_for_different_futures} illustrates how they can influence different system states over time, while \Cref{subsubsec:methodology_choices_in_decision_frameworks} discusses various modelling options that allow to incorporate them into system transformation planning.

\subsubsection{Transformation pathways for different futures}
\label{subsubsec:transformation_pathways_for_different_futures}

\begin{figure*}[h]
    \centering
    \includegraphics[width=\textwidth]{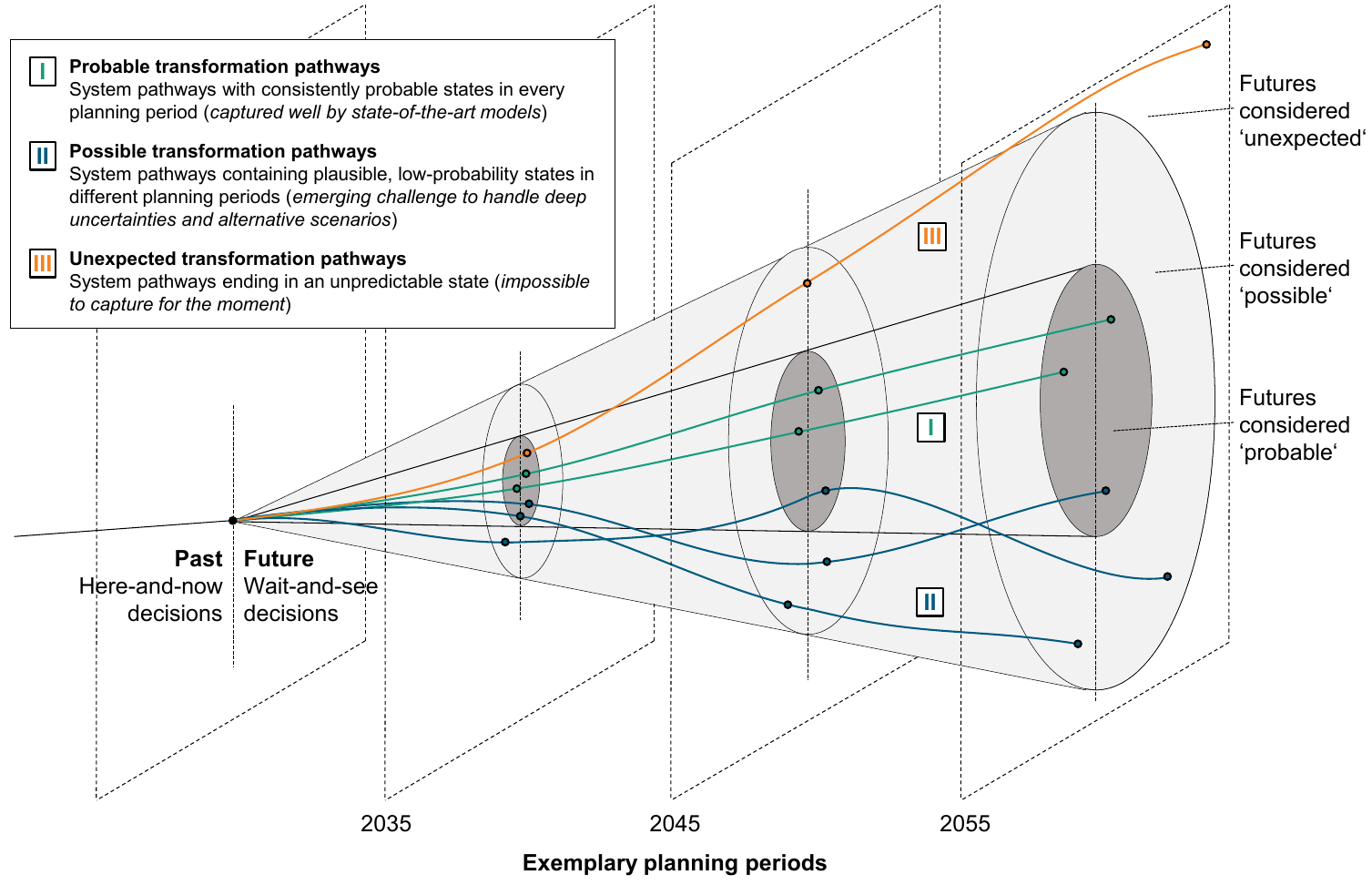}\\
    \caption{Potential transformation pathways along the ``futures cone'', own illustration inspired by and adapted from 
    \cite{Voros.2003}, \cite{vanDorsser.2018}, and \cite{McCollum.2020}. Note that this is a schematic three-dimensional representation, whereas reality is multidimensional. Furthermore, the shown system pathway trajectories are non-exhaustive as other viable alternatives may exist.}
    \label{fig_cone_transition_pathway} 
\end{figure*}

Inspired by the ``futures cone'' introduced by \cite{Voros.2003}, \cite{vanDorsser.2018}, and \cite{McCollum.2020}, \Cref{fig_cone_transition_pathway} illustrates the potential evolution of different energy system states across several years.
Time is represented along the horizontal axis, exemplified by three future planning periods at 10-year intervals.
The inner cone symbolises the set of system states that are considered `probable', while the outer cone symbolises the set of system states that are considered less probable but still `possible'.
The inner cone symbolises system states classified as `probable' while the outer cone represents less likely but still `possible' states.
System states beyond these cones are currently unpredictable and therefore labelled as `unexpected'.
A `system state' may encompass numerous variables, including installed capacities, energy demands, technological costs, energy prices, emission budgets, or external factors such as climate conditions.

\Cref{fig_cone_transition_pathway} demonstrates various energy system transformation pathways, grouped into three categories:

\begin{enumerate}[label=\Roman*.]
    \item \textbf{Probable transformation pathways} consistently contain probable system states across all planning periods and are readily captured by current state-of-the-art energy system models.
    \item \textbf{Possible transformation pathways} contain plausible, low-probability system states in different planning periods, posing an emerging challenge to handle deep uncertainties and alternative scenarios in energy system planning.
    \item \textbf{Unexpected transformation pathways} culminate in currently unpredictable system states not currently manageable within existing modelling frameworks.
\end{enumerate}

Addressing these future modelling challenges requires capturing an increasingly broad range of potential system developments within the outer cone.
Achieving this may necessitate adopting innovative and scalable modelling methodologies beyond those traditionally employed.

\subsubsection{Methodology choices in decision frameworks}
\label{subsubsec:methodology_choices_in_decision_frameworks}

Resilient energy system planning requires decision-making frameworks capable of addressing uncertainties, including shock events and \glspl{SBP}.
These frameworks generally fall into qualitative and quantitative approaches.
Qualitative approaches emphasise conceptual understanding and stakeholder engagement to identify risks and formulate strategies.
In contrast, quantitative approaches employ analytical methods, such as simulation and optimisation, to derive actionable insights.
\par
Simulation methods are primarily predictive and aim to replicate system behaviour under various predefined scenarios.
They are widely employed to assess supply resilience and security by modelling energy systems' performance under different stress conditions.
However, simulation alone provides limited prescriptive guidance for decision-making.
\par
Optimisation methods, on the other hand, are prescriptive and focus on identifying the best course of action under given constraints and objectives.
These methods are particularly suited to planning and managing energy systems under uncertainty.
Decision support frameworks can achieve a balanced trade-off between competing priorities by integrating system performance indicators, e.g. costs, technical reliability, and emissions, into the optimisation objective.
\par
Optimisation under uncertainty provides an alternative to explicitly account for shock events and \glspl{SBP} and the available recourse options when designing transformation pathways towards clean energy systems.
Depending on the chosen risk metrics incorporated into objective functions or constraints, several approaches can be employed, including risk-neutral and risk-averse stochastic optimisation, robust optimisation, distributionally robust optimisation, and chance-constrained optimisation~\citep{Roald.2023}.
These methodologies differ in their representation of uncertainty or ambiguity sets, how uncertainties propagate through the decision model, and their complexity in terms of implementation and computational requirements.
Recent examples illustrating the use of stochastic optimisation in energy system modelling include case studies employing EMPIRE \citep{Ahang.2025,Backe.2022} and the EMPRISE framework \citep{Schmitz.2024,Frischmuth.2024}.

\subsection{Further challenges in resilience assessment}
\label{subsec:challenges_in_the_assessment_of_resilience}

Evaluating resilience in integrated energy systems often reveals inherent contradictions, as certain resilience-enhancing solutions may simultaneously pose new challenges.
Two notable challenges are briefly outlined below.

\subsubsection{Different system performances}

\Cref{subsec:impact_shock_SBP_over_time} has demonstrated that various performance dimensions of energy systems --- technical, economic, and environmental --- can conflict.
A critical question emerges for incidents that adversely impact all three system performances: Is there a hierarchy among these dimensions?
Specifically, does the recovery of one performance dimension take precedence over the others?
Moreover, the interplay between these aspects remains ambiguous.
How should technical, economic, and environmental performances be balanced or converted to achieve optimal outcomes?
When assessing overall system performance, pivotal decisions must be made: Should each dimension be optimised individually, or should an integrated, holistic approach be adopted to achieve a balanced outcome?
Deciding whether to optimise each dimension separately or adopt an integrated approach to balance all performance areas is crucial for resilient and sustainable energy system transformations.

\subsubsection{Integrated commodities}
Integrated commodities refer to energy resources and technologies that are interconnected and coordinated to optimise energy production, distribution, and consumption.
Key examples include natural gas, hydrogen, and electricity, all pivotal to modern energy systems.
The interaction among these commodities offers both opportunities and challenges.
On the one hand, hybridisation fosters flexibility across sectors.
For instance, as electrification advances, electricity can address conventional demand in the electricity sector, power heat pumps in the heating sector, and serve as an energy source for electric vehicles in the transportation sector.
Furthermore, integration enables cross-sectoral energy flows, where one commodity can supplement another, enhancing overall system adaptability (see \Cref{subsec:adaptive_recourse_options}).
\par
On the other hand, this interconnectedness also introduces vulnerabilities.
Periods of low renewable generation, such as extended low-wind and low-solar events (known as ``Dunkelflaute''), can lead to substantial disruptions in electricity supply, potentially cascading across dependent sectors.
Heating systems relying on electric heat pumps, electricity-driven industrial processes, and electrified transportation networks become vulnerable, exacerbating system risks.
\par
Therefore, integrating commodities effectively with the appropriate recourse options (recall \Cref{subsec:absorptive_recourse_options,subsec:adaptive_recourse_options,subsec:transformative_recourse_options}), despite its complexity, remains fundamental for resilient resilient system development planning.

\subsection{System resilience-cost trade-offs}
\label{sec:trade_off}
A central challenge for policymakers and system planners is understanding and quantifying the trade-offs between resilience levels and associated system costs.
Clarifying these resilience-cost trade-offs is critical to making informed decisions about acceptable expenditure levels for enhanced system resilience.
\par
\begin{figure}
    \centering
    \begin{tikzpicture}
\begin{axis}[xmin=0,xmax=10,ymin=0,ymax=100,axis lines = none,domain=0:10,samples=50]

\definecolor{fhgreen}{RGB}{23,156,125}
\definecolor{fhblue}{RGB}{0,91,127}
\definecolor{fhgrey}{RGB}{166,187,200}
\definecolor{fhorange}{RGB}{245,130,32}
\definecolor{fhpurple}{RGB}{124,21,77}
\definecolor{fhteal}{RGB}{0,133,152}
\definecolor{fhtrq}{RGB}{57,193,205}

\addplot[fhgreen, thick, domain=0:4.5555] {x < 1 ? 2 : 2 * exp(1.1*(x-1))}; 

\addplot[fhblue, thick, domain=0:10] {x < 3 ? 2 : 2 * exp(0.35*(x-3))};

\addplot[fhgrey, const plot mark mid, domain=0:10] coordinates {(0,2) (3,3) (4,12) (5,16) (6,22) (8,30) (9,38) (10,50)};
\addplot[fhtrq, domain=0:10] coordinates {(0,2) (4,3) (4,5) (5,6) (5,8) (6,10) (6,12) (8,18) (8,22) (9,27) (9,32) (10,38) (10,42)};

\addplot[fhorange, thick, domain=2:10] {2 + 50/(1 + exp(-3*(x - 4.5)))}; 



\draw[dotted, thick, fhpurple] (4.0,0) -- (4.0,100);

\draw[dotted, thick, fhteal] (0,40) -- (10,40);

\draw[-{Latex[open]}, thick, color=fhpurple] (4.0, 87.5) -- (5.0, 87.5);

\draw[-{Latex[open]}, thick, color=fhteal] (1.75, 40) -- (1.75, 30);

\node[align=center, draw=fhpurple, fill=white, text=black] at (axis cs:2.1,87.5) {Desired system\\resilience level};
\node[align=center, draw=fhteal, fill=white, text=black] at (axis cs:1.75,50) {Acceptable\\system costs};

\draw[-{Circle[open]}, thick, color=black] (8, 80) -- (4.1,70); 
\draw[-{Circle[open]}, thick, color=black] (8, 80) -- (5.7,50); 
\draw[-{Circle[open]}, thick, color=black] (8, 80) -- (9,15); 
\draw[-{Circle[open]}, thick, color=black] (8, 80) -- (6,21);
\draw[-{Circle[open]}, thick, color=black] (8, 80) -- (9.55,34);
\node[align=center, draw=black, fill=white, text=black] at (axis cs:8,80) {Shape of\\individual or\\ aggregate\\trade-off curves};

\end{axis}

\begin{axis}[
    axis lines=middle,
    axis on top=true,
    xlabel={\textbf{System resilience level}},
    ylabel={\textbf{System costs}},
    xlabel style={at={(axis description cs:0.5,-0.03)}, anchor=north},
    ylabel style={at={(axis description cs:-0.03,0.5)}, anchor=south, rotate=90},
    xmin=0, xmax=10,
    ymin=0, ymax=100,
    xtick=\empty,
    ytick=\empty,
    grid=major
    ]
\end{axis}
\end{tikzpicture}
    \caption{Schematic illustration of the system resilience vs. system costs trade-off, own illustration.}
    \label{fig_tradeoff_curves}
\end{figure}
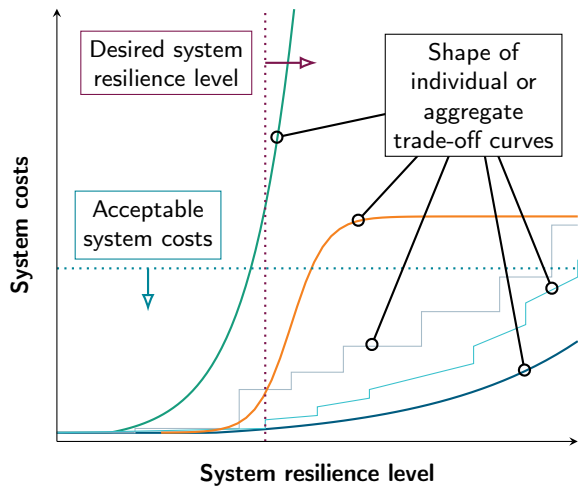

\Cref{fig_tradeoff_curves} illustrates schematic examples of trade-offs between system resilience and system costs, highlighting that these relationships are complex and highly context-dependent.
For illustrative purposes, various trade-off shapes are depicted, including exponential (green), S-shaped (orange), stepwise (grey), slow incremental (dark blue), and aggregate (light blue):
\begin{itemize}
\item \textbf{Exponential cost increase}: Here, costs rise sharply as resilience approaches its maximum, indicating that initial improvements are relatively affordable, whereas achieving near-maximum resilience becomes disproportionately costly.
\item \textbf{S-shaped (logistic) curve}: Initial resilience improvements require substantial effort with moderate costs. However, costs plateau after reaching a saturation point, reflecting diminishing returns beyond a certain resilience level.
\item \textbf{Stepwise increase}: Trade-off curves illustrate abrupt cost increases at specific resilience thresholds, likely corresponding to significant infrastructure investments or policy-driven decision points.
\item \textbf{Slow incremental increase}: In this scenario, resilience improvements lead to gradual and relatively steady cost increases, suggesting consistent returns on resilience investments.
\end{itemize}
\par
Understanding the factors that influence these different trade-off shapes is essential.
Important questions include: How does the event type affect the cost-resilience relationship? How does the choice of absorptive, adaptive, or transformative resilience capacities impact the shape and steepness of the trade-off curve?
\par
A further critical aspect is determining political and societal acceptance of different resilience levels and associated costs.
Policymakers must address what level of resilience is desirable or acceptable for each specific shock event or \gls{SBP}, and what level of additional system cost is justified or politically feasible to achieve this resilience.
\par
Beyond individual disturbances and recourse actions, there is a critical need to understand potential aggregated trade-off curves.
For instance, are there synergistic effects across multiple resilience measures?
Are different resilience-enhancing actions complementary, creating efficiencies, or are they independent or even counterproductive?
\par
Addressing these complex questions requires more explicit political guidelines and a better quantitative understanding of how resilience and costs interact. 

\section{Conclusions and policy implications}
\label{sec:conclusion}

Policymakers face the critical task of balancing immediate energy demands with long-term sustainability objectives.
Effective policies must be adaptable to unknown and uncertain futures and ensure resilience amidst evolving energy systems \citep{vanDorsser.2018}.
\par
Reviewing critical energy security and resilience concepts, we emphasise their importance in light of recent global events such as the pandemic, geopolitical conflicts, supply chain disruptions, and climate change impacts.
The study highlights the need for a thorough understanding of these concepts to support the robustness of future integrated energy systems. Key contributions of this work include:
\begin{enumerate}[label=\Roman*.]
    \item \textbf{Conceptual clarification}: We provide a detailed review of existing definitions and interrelations between energy security, resilience, and related terms, enhancing clarity for policymakers and system planners to develop effective strategies.
    \item \textbf{Classification of shock events and \glspl{SBP}}: We categorise system disturbances into shock events and \glspl{SBP}, emphasising that shock events arise suddenly and unexpectedly, while \glspl{SBP} evolve gradually, escalating in urgency over time. By providing examples, we demonstrate their distinct impacts on technical, economic, and environmental system performance, thereby enhancing the understanding of the unique challenges posed by these disturbances.
    \item \textbf{Recourse options}: We compile relevant recourse options across resilience capacity levels and system planning horizons. These options provide actionable strategies for enhancing the resilience of integrated energy systems.
    \item \textbf{Recommendations for energy system modelling}: Many studies still rely on deterministic planning approaches that do not account for potential threats and hazards. We recommend more frequent integration of shock events and \glspl{SBP} into future energy system planning to enhance resilience. Incorporating these uncertainties enables forward-looking decision-making, helping to mitigate potential disruptions and strengthen resilience in future energy systems.
    \item \textbf{Clarifying resilience-cost trade-offs}: We highlight the necessity for policymakers to better understand and quantify individual and aggregate trade-offs between resilience levels and associated costs. We recognise the need for clearer political guidelines regarding desired resilience levels and acceptable additional costs. Policymakers must define how much additional expenditure is justified to achieve a more resilient system.
\end{enumerate}
Our findings emphasise that incorporating threats and hazardous events into energy system planning is essential for resilience and avoiding high recourse costs.
Policymakers should, therefore, ensure that their decisions are informed by energy system planning tools that effectively assess future uncertainties and their impact on energy security and resilience.
Achieving this requires a mutual understanding: modellers need clear political guidelines, while policymakers depend on robust modelling analyses for informed decision making.
Existing energy-only markets may not sufficiently incentivise for preparation against shock events or \glspl{SBP}. 
Hence, additional mechanisms, such as capacity markets or targeted subsidies, should be explored to encourage adequate investment in resilience capacities and recourse options.
As a critical next step, quantifying the desired resilience levels and acceptable cost thresholds is essential for informed, strategic planning.

\printcredits


\section*{Declaration of generative AI and AI-assisted technologies in the writing process}
During the preparation of this work, the authors used generative AI to improve language and readability. After using these tools, the authors reviewed and edited the content as needed and take full responsibility for the publication's content.

\section*{Declaration of competing interest}
The authors declare that they have no known competing financial interests or personal relationships that could have appeared to influence the work reported in this paper.

\section*{Funding}
The work of this paper has been performed as part of the project \href{https://www.iee.fraunhofer.de/rewards}{\textit{REWARDS}}, which receives funding from the German Federal Ministry for Economic Affairs and Climate Action (BMWK) under funding reference numbers 03EI1083A, 03EI1083B, 03EI1083C, 03EI1083D.

\section*{Acknowledgments}
The authors would like to thank Valentin Bertsch and Magnus Korp{\aa}s for valuable comments and interesting discussions.

\bibliographystyle{elsarticle-harv}

\bibliography{cas-refs}

\appendix

\section{Further related terms}\label{app:further_terms}

During the literature review, additional terms emerged that require definition and categorisation.
While not the primary focus of this study, the following sections on efficiency, vulnerability and robustness provide contextual insights.

\subsection{Efficiency}
The efficiency of an energy system is commonly defined as the ratio of energy demand to actual energy use \citep{Kardas.2023}.
In the literature, efficiency is frequently referred to as one of several dimensions of energy security (e.g. in \citealp{Ang.2015,Esfahani.2021,Fouladvand.2024}).
However, in \cite{Kardas.2023}, it is categorised as a dimension of energy sovereignty.
Regarding energy system's resilience, some studies (e.g. \citealp{Zhou.2023,Brand.2016}) suggest that efficiency and resilience may be at odds with each other.

\subsection{Vulnerability}
Vulnerability generally refers to the consequences of hazardous events \citep{Izadi.2021}.
More specifically, energy vulnerability is defined as ``the degree to which an energy system is unable to cope with selected adverse events and risks to fall into traps in economic, social, environmental and institutional terms'' \citep{Gatto.2020,Gatto.2020c}.
\cite{Mohanty.2024} further classify energy system vulnerability into physical vulnerability, cyber vulnerability, and cyber-physical vulnerability.
As noted in \cite{Aldieri.2021}, energy efficiency and resilience are expected to mitigate energy vulnerability.

\subsection{Robustness}
Robustness is defined as the ability of components to resist external and internal influences, ensuring that its structure and functionality remain intact during system during normal operation \citep{VDE.2020}.
The definition provided in \cite{Zhou.2023} expands on this definition, describing robustness as a ``flexibility that enables the system to function properly even when the structure is broken or damaged''.
There is a consensus in the literature that robustness is not equal to resilience and vice versa \citep{Hanke.2021}.
Instead, \cite{Bitkom.2018} and \cite{Braun.2020} argue that resilience is an overarching concept that extends beyond robustness.
This distinction is further emphasised by the \cite{NIAC.2010}, which identifies robustness as just one of several criteria contributing to resilience.

\end{document}